\documentclass[nofootinbib,reprint,floatfix, longbibliography, superscriptaddress]{revtex4-1}
\usepackage{amsmath}
\usepackage{amssymb}
\usepackage{graphicx}
\usepackage{verbatim}
\usepackage{color}
\usepackage{subfigure}
\usepackage{appendix}
\usepackage[pdftex, breaklinks,colorlinks, citecolor=blue, urlcolor=blue]{hyperref}

\begin{document}

\title{Bidirectional and efficient conversion between microwave and optical light}
\author{R. W. Andrews}
\affiliation{JILA, University of Colorado and NIST, Boulder, Colorado 80309, USA}
\affiliation{Department of Physics, University of Colorado, Boulder, Colorado 80309, USA}

\author{R. W. Peterson}
\affiliation{JILA, University of Colorado and NIST, Boulder, Colorado 80309, USA}
\affiliation{Department of Physics, University of Colorado, Boulder, Colorado 80309, USA}

\author{T. P. Purdy}
\affiliation{JILA, University of Colorado and NIST, Boulder, Colorado 80309, USA}
\affiliation{Department of Physics, University of Colorado, Boulder, Colorado 80309, USA}

\author{K. Cicak}
\affiliation{National Institute of Standards and Technology (NIST), Boulder, Colorado 80305, USA}

\author{R. W. Simmonds}
\affiliation{National Institute of Standards and Technology (NIST), Boulder, Colorado 80305, USA}

\author{C. A. Regal}
\affiliation{JILA, University of Colorado and NIST, Boulder, Colorado 80309, USA}
\affiliation{Department of Physics, University of Colorado, Boulder, Colorado 80309, USA}

\author{K. W. Lehnert}
\affiliation{JILA, University of Colorado and NIST, Boulder, Colorado 80309, USA}
\affiliation{Department of Physics, University of Colorado, Boulder, Colorado 80309, USA}
\affiliation{National Institute of Standards and Technology (NIST), Boulder, Colorado 80305, USA}

\begin{abstract}

Converting low-frequency electrical signals into much higher frequency optical signals has enabled modern communications networks to leverage both the strengths of microfabricated electrical circuits and optical fiber transmission, allowing information networks to grow in size and complexity.  
A microwave-to-optical converter in a quantum information network could provide similar gains by linking quantum processors via low-loss optical fibers and enabling a large-scale quantum network. However, no current technology can convert low-frequency microwave signals into high-frequency optical signals while preserving their fragile quantum state.  For this demanding application, a converter must provide a near-unitary transformation between different frequencies; that is, the ideal transformation is reversible, coherent, and lossless.  Here we demonstrate a converter that reversibly, coherently, and efficiently links the microwave and optical portions of the electromagnetic spectrum.  We use our converter to transfer classical signals between microwave and optical light with conversion efficiencies of $\sim$10\%, and achieve performance sufficient to transfer quantum states if the device were further precooled from its current 4 kelvin operating temperature to below 40 millikelvin. The converter uses a mechanically compliant membrane to interface optical light with superconducting microwave circuitry, and this unique combination of technologies may provide a way to link distant nodes of a quantum information network.

\end{abstract}
\maketitle

\subsection*{Introduction}

Modern communication networks manipulate information at several gigahertz with microprocessors and distribute information at hundreds of terahertz via optical fibers.  A similar frequency dichotomy is developing in quantum information processing.  Superconducting qubits operating at several gigahertz have recently emerged as promising high-fidelity and intrinsically scalable quantum processors \cite{clarke2008, girvin2008, schoelkopf2013}.  Conversely, optical frequencies provide access to low-loss transmission \cite{jelena2009} and long-lived quantum-compatible storage \cite{nori2011,langer2005}.  Converting information between gigahertz-frequency ``microwave light'' that can be deftly manipulated and terahertz-frequency ``optical light'' that can be efficiently distributed will enable small-scale quantum systems \cite{rempe2012,schoelkopf2012, lucero2012} to be combined into larger, fully-functional quantum networks \cite{kimble2008, obrien2010}.  But no current technology can transform information between these vastly different frequencies while preserving the fragile quantum state of the information.  For this demanding application, a frequency converter must provide a near-unitary transformation between microwave light and optical light; that is, the ideal transformation is reversible, coherent, and lossless.

Certain nonlinear materials provide a link between microwave and optical light, and these are commonly used in electro-optic modulators (EOMs) for just this purpose.  While EOMs  might be capable of reversible frequency conversion \cite{tsang2010,tsang2011}, such conversion has not yet been demonstrated, and even optimized EOMs \cite{levi2001,maleki2003} have predicted photon number efficiencies of only a few $10^{-4}$ \cite{maleki2003,tsang2010, tsang2011}.  Other intermediate objects that interact with both microwave and optical light can be used to create the nonlinearity necessary for frequency conversion.  Proposed intermediaries include clouds of ultracold atoms \cite{taylor2012, schied2009}, ensembles of spins \cite{atac2009, marcos2010}, and mechanical resonators \cite{naeini2011, Lehnert2011, cleland2013}.  All converters face the challenge of integrating optical light with the cryogenic temperatures needed for low-noise microwave signals and superconducting circuitry.  Here, we demonstrate a cryogenic converter that incorporates a mechanical resonator with optical light and superconducting circuitry, and use it to reversibly transform classical signals between microwave and optical frequencies.


Early experiments used microwave light \cite{braginsky1970, gozzini1985} and optical light \cite{dorsel1983} to manipulate mechanical resonators and study the interaction between light and a vibrating mass \cite{braginsky1967, caves1980}. The fields of electromechanics and optomechanics have since evolved at a remarkably commensurate pace.  Both microwave and optical light have been separately used to cool a mechanical resonator to its quantum ground state of motion \cite{jdteufel2011, chan2011}.  This same interaction allows the mechanical resonator to serve as an information storage medium \cite{verhagen2012,palomaki2013}, and opens up the possibility of high-fidelity frequency conversion \cite{tian2010, wang2012, tian2012,vitali2012,mcgee2013}.  By combining the technologies of electromechanics and optomechanics, we simultaneously couple a mechanical resonator to both a microwave circuit and an optical cavity.  This simultaneous coupling allows information in the microwave domain to stream through the mechanical resonator and emerge as optical light, and vice-versa.  We show that the transformation is reversible and coherent, and further demonstrate a photon number efficiency of 0.086$\pm$.007 and a transfer bandwidth of 30 kHz.  Furthermore, the performance is sufficient for nearly noiseless frequency conversion if we can further precool the device from its current 4 kelvin operating temperature to dilution refrigerator temperatures below 40 millikelvin.  


\subsection*{Cavity electro-opto-mechanics}

\begin{figure}
\begin{minipage}{\linewidth}
\scalebox{.6}{\includegraphics{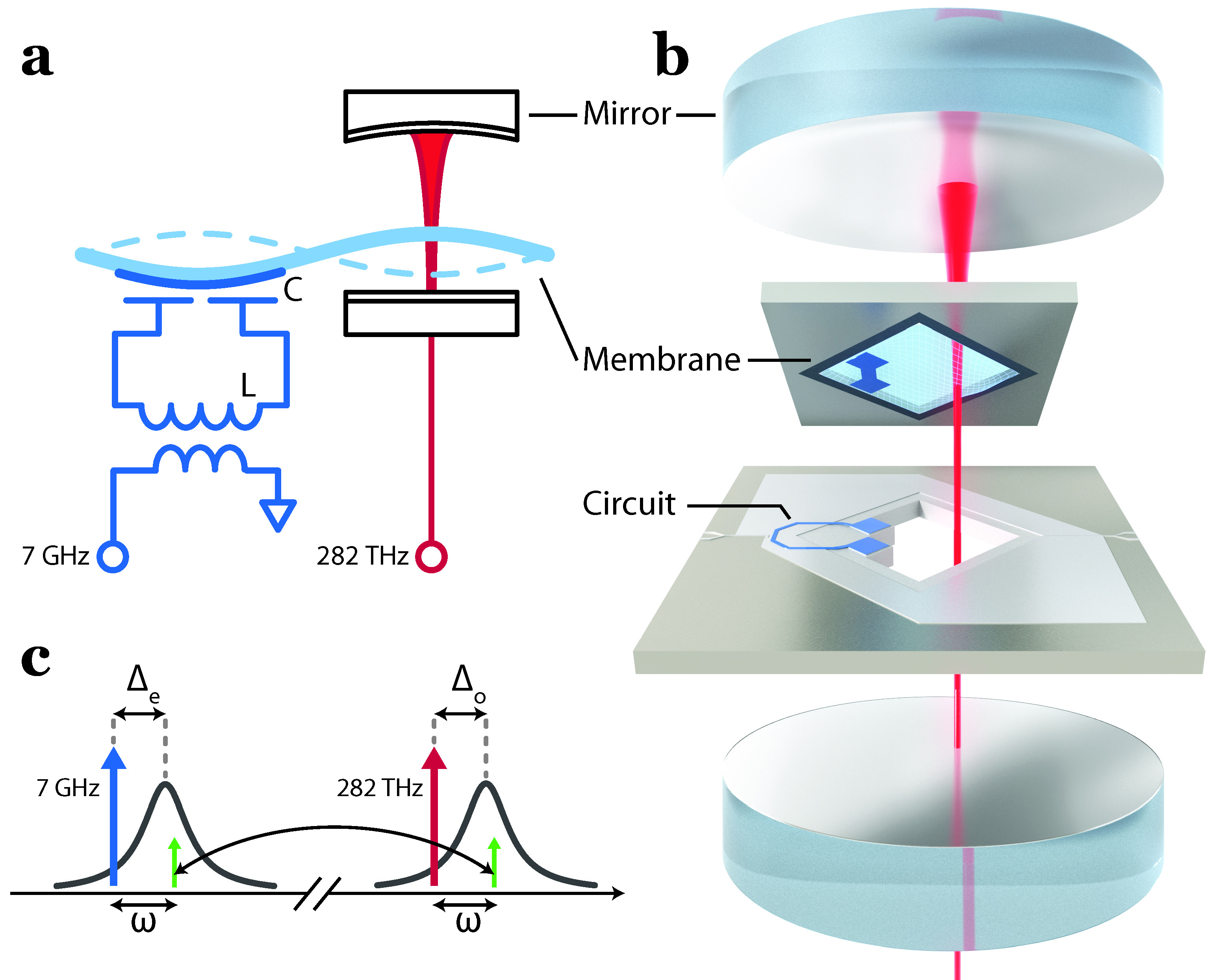}}
\caption{Layout and operation of microwave-to-optical converter ($\mathbf{a}$) A stoichiometric silicon nitride (Si$_3$N$_4$) membrane (light blue) that has been partially covered with niobium (dark blue) interacts with an inductor-capacitor (LC) circuit that forms the microwave resonator and a Fabry-Perot cavity that forms the optical resonator (mode shown in red).  Propagating light fields are coupled to the microwave resonator with an inductive coupler, and to the optical resonator with a slightly transmissive input mirror.  ($\mathbf{b}$) The membrane is suspended within a silicon frame, and the microwave circuitry is lithographically patterned on a separate silicon substrate.  The two silicon chips are brought together so that the niobium-covered portion of the membrane comes to within 500 nanometers of the microwave circuitry, thus forming the electromechanical system.  The electromechanical system is then placed inside the optical resonator.  The entire structure is made to be cryogenically compatible.  ($\mathbf{c}$) A frequency domain representation of the conversion process.  A strong microwave pump (blue arrow) is applied below the microwave resonance (response shown as a black curve) with detuning $\Delta_{\mathrm{e}}$.  Likewise, a strong optical pump (red arrow) is applied below the optical resonance (response shown as a black curve) with detuning $\Delta_{\mathrm{o}}$.  This allows a signal (green arrow) to be up- or downconverted in frequency.}
\label{schematic1}
\end{minipage}
\end{figure}

Our converter consists of two electromagnetic resonators, one at an optical frequency and one at a microwave frequency, that share a mechanical resonator.  The mechanical resonator is formed by a thin silicon nitride membrane that is free to vibrate (Fig. \ref{schematic1}).  The optical frequency resonator consists of a Fabry-Perot cavity, and as the membrane vibrates it moves along the optical intensity standing wave and modulates the resonant frequency of the optical cavity \cite{thompson2008, purdy2013}.  The membrane is partially coated with a thin layer of niobium (which superconducts at temperatures below 9 kelvin), and this electrically conductive portion is part of a capacitor in an inductor-capacitor circuit that forms the microwave resonator (K.C., {\it{in prep.}})\cite{yu2012,polzik2013}.  As the membrane vibrates, it modulates the capacitance of the microwave circuit, and thus its resonant frequency.  Even though the electromagnetic resonators are at vastly different frequencies (7 GHz and 282 THz), the coupling mechanism is equivalent: A nanometer of membrane motion shifts the microwave resonant frequency by approximately 4 MHz, and shifts the optical resonant frequency by approximately 40 MHz, giving coupling constants of $G_e \approx 4$ MHz/nm and $G_o \approx 40$ MHz/nm.

During the experiment, a strong pump tone is applied below the resonant frequency of each electromagnetic resonator.  The pumps enhance the electromechanical and optomechanical interaction, and the mechanical resonator exchanges information with the microwave and optical resonators at rates $g_{\mathrm{e}}=G_{\mathrm{e}}x_{\mathrm{zp}}\sqrt{n_{\mathrm{e}}}$ and $g_{\mathrm{o}}=G_{\mathrm{o}}x_{\mathrm{zp}}\sqrt{n_{\mathrm{o}}}$, respectively, where $x_{\mathrm{zp}}$ is the zero-point motion of the mechanical resonator and $n_\mathrm{e}$ ($n_\mathrm{o}$) is the number of photons in the microwave (optical) resonator induced by the microwave (optical) pump.  The expressions for $g_{\mathrm{e}}$ and $g_{\mathrm{o}}$ take on this simple form in the resolved-sideband limit (defined as $4\omega_{\mathrm{m}} \gg \kappa_{\mathrm{e}},\kappa_{\mathrm{o}}$, where $\omega_{\mathrm{m}}$ is the frequency of the vibrational mode of the mechanical resonator and $\kappa_{\mathrm{e}}$ and $\kappa_{\mathrm{o}}$ are the energy decay rates of the microwave and optical resonators, respectively); however, these coupling rates can always be independently adjusted in-situ by changing the strength of the pumps and altering $n_{\mathrm{e}}$ and $n_{\mathrm{o}}$ \cite{palomaki2013}.  This coherent exchange of information between electromagnetic and vibrational modes is capable of quantum state preserving frequency conversion \cite{akram2010, zhang2003}.  

 A full description of the system includes the inputs and outputs of the microwave, optical, and mechanical resonators.  Of all the energy leaving the microwave (optical) resonator, only a fraction $\eta_{\mathrm{e}}$ ($\eta_{\mathrm{o}}$) exits into the propagating mode that we collect.  Some energy is absorbed in the resonators themselves, and the optical resonator emits light into a particular spatial mode that does not perfectly match the spatial mode of the incident light field.   The fraction of light we collect can be expressed as $\eta_{\mathrm{e}}=\kappa_{\mathrm{e,ext}}/\kappa_{\mathrm{e}}$ and $\eta_{\mathrm{o}}=\epsilon \kappa_{\mathrm{o,ext}}/\kappa_{\mathrm{o}}$, where $\kappa_{\mathrm{e,ext}}$ ($\kappa_{\mathrm{o,ext}}$) is the rate at which energy leaves the microwave (optical) resonator into propagating fields, and $\epsilon$ is the optical mode matching.  If $\kappa_{\mathrm{e}} \gg g_{\mathrm{e}}$ and $\kappa_{\mathrm{o}}\gg g_{\mathrm{o}}$, the electromagnetic resonators couple energy and information in freely propagating microwave (optical) modes to a vibrational mode of the mechanical resonator at a rate $\Gamma_{\mathrm{e}}$ ($\Gamma_{\mathrm{o}}$), which has the simple form $\Gamma_{\mathrm{e}}=4 g_{\mathrm{e}}^{2}/\kappa_{\mathrm{e}}$ ($\Gamma_{\mathrm{o}}=4 g_{\mathrm{o}}^{2}/\kappa_{\mathrm{o}}$) in the resolved-sideband limit \cite{hill2012}.

\begin{figure}
\begin{minipage}{\linewidth}
\scalebox{.7}{\includegraphics{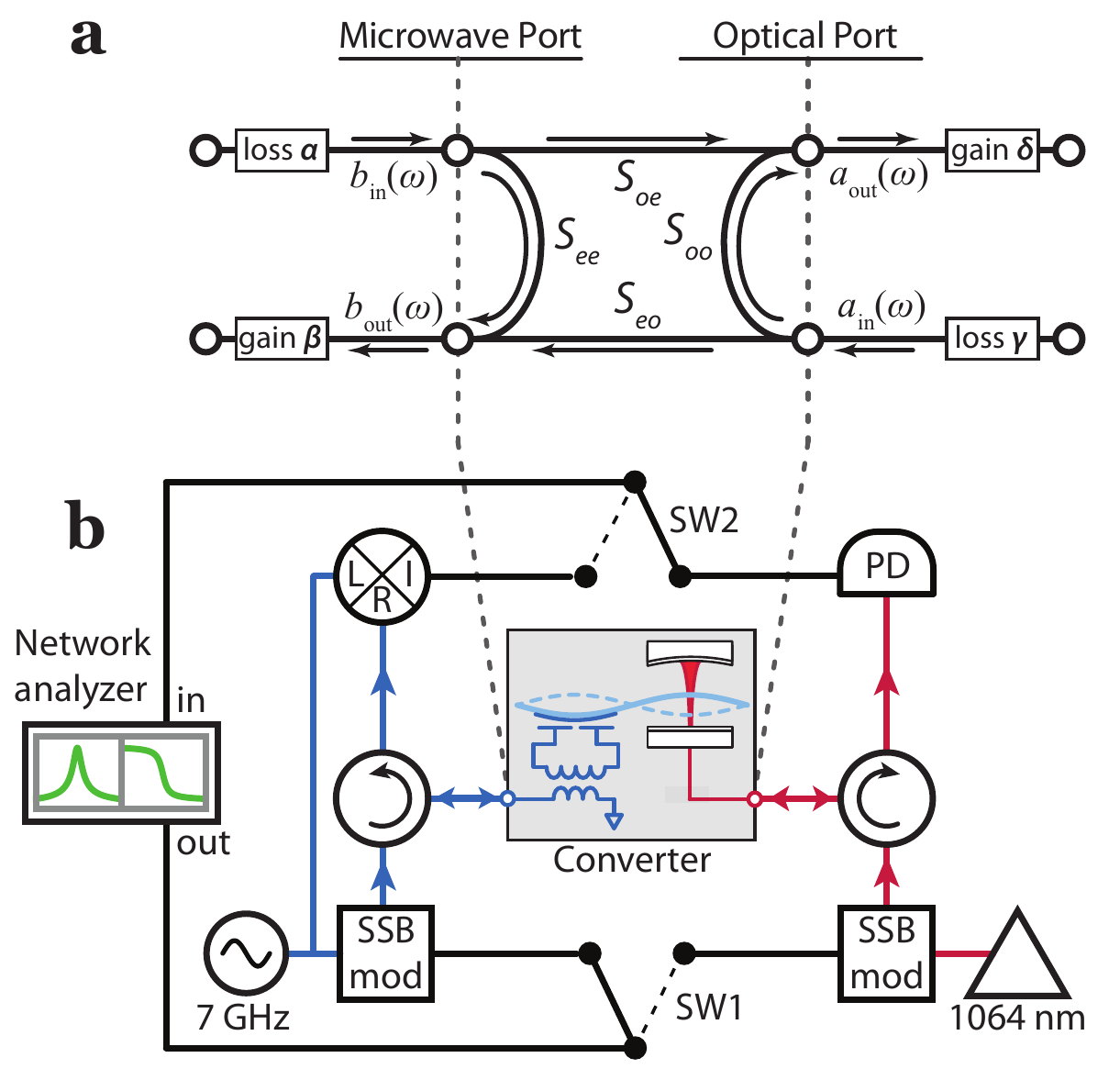}}
\caption{Measurement network.  ($\mathbf{a}$) Converter $S$-parameters are defined between the ports of the microwave and optical resonators.  To isolate the converter performance from the gains and losses of other components used in our measurement, we measure the forward and reverse transmission, $\alpha S_{\mathrm{oe}} \delta$ and $\gamma S_{\mathrm{eo}}\beta$, and the off-resonant reflections, $\alpha\beta$ and $\gamma\delta$.   This gives $S_{\mathrm{oe}}S_{\mathrm{eo}}=(\alpha S_{\mathrm{oe}}\delta)(\gamma S_{\mathrm{eo}}\beta)/(\alpha\beta)(\gamma\delta)$, a quantity insensitive to imperfect measurement of the separate gains and losses $\alpha$, $\beta$, $\gamma$, $\delta$.  The off-resonant reflections are measured by tuning the injected signal away from the resonator's central frequency, where the converter acts as a near-perfect mirror. ($\mathbf{b}$) A vector network analyzer generates a signal near the mechanical resonant frequency, $\omega_{\mathrm{m}}$.  This signal is mixed to a higher frequency using either the microwave pump or the optical pump as a carrier via single-sideband (SSB) modulation.  After interacting with the converter, the outgoing fields are separated using circulators, and detected with homodyne detection (for microwave frequencies) or direct detection with a photodiode (for optical frequencies).   The detected signal is returned to the vector network analyzer, where it is referenced to the original signal to obtain a magnitude and phase shift.  }
\label{setup}
\end{minipage}
\end{figure}

During upconversion, an injected microwave field enters the converter at a frequency $\omega$ above the microwave pump, enters and exits the mechanical resonator as determined by coupling rates $\Gamma_{\mathrm{e}}$ and $\Gamma_{\mathrm{o}}$, and emerges as an outgoing optical field at a frequency $\omega$ above the optical pump.   A frequency-domain representation of the process is shown in Fig. \ref{schematic1}c.  Converter performance is characterized by how efficiently the input microwave field, ${b}_{\mathrm{in}}(\omega)$, is transformed into an output optical field, ${a}_{\mathrm{out}}(\omega)$, and vice versa.  The ratio ${a}_{\mathrm{out}}(\omega)/{b}_{\mathrm{in}}(\omega)\equiv S_{\mathrm{oe}}(\omega)$ is one of four scattering parameters that characterize the two-port network formed by the converter (Fig. \ref{setup}a).  The fields ${b}_{\mathrm{in}}(\omega)$ and ${a}_{\mathrm{out}}(\omega)$ have units of  $(\mathrm{number\cdot sec})^{1/2}$, and so the apparent photon number efficiency for upconversion is given by $|{a}_{\mathrm{out}}(\omega)/{b}_{\mathrm{in}}(\omega)|^{2}=|S_{\mathrm{oe}}(\omega)|^{2}$.

During the experiment, the converter is integrated into a larger network.  To predict and measure converter performance, we need to specify which components are part of the converter, and which are part of the measurement network.  We choose to define the converter as all the components between the inductive coupler of the microwave resonator and the input mirror of the optical resonator (Fig. \ref{setup}).  Converter performance then includes internal losses in the microwave, optical, and mechanical resonators and imperfect optical mode matching, but excludes losses and gains in other components that are used in our measurement.  With this definition, the converter is a stand-alone component that can be readily integrated into other networks.  We can predict converter efficiency using a Hamiltonian that includes radiation pressure coupling \cite{law1995} to generate Heisenberg-Langevin equations of motion (see S.I.).  This analysis predicts
\begin{align}
\label{trans}
S_{\mathrm{oe}}(\omega)= \frac{\sqrt{\Gamma_{\mathrm{e}}\Gamma_{\mathrm{o}}}}{-\imath\left( \omega-\omega_{\mathrm{m}} \right)+ \left( \Gamma_{\mathrm{e}}+\Gamma_{\mathrm{o}}+\kappa_{\mathrm{m}} \right)/2} \times \sqrt{\mathcal{A}\eta_{\mathrm{e}}\eta_{\mathrm{o}}}
\end{align}
where $\kappa_{\mathrm{m}}$ is the intrinsic mechanical damping and $\mathcal{A}=\mathcal{A}_{\mathrm{e}}\mathcal{A}_{\mathrm{o}}$ is conversion gain, with $\mathcal{A}_{\mathrm{e}}=1+(\kappa_{\mathrm{e}}/4\omega_{\mathrm{m}})^2$ and $\mathcal{A}_{\mathrm{o}}=1+(\kappa_{\mathrm{o}}/4\omega_{\mathrm{m}})^2$.  The reverse process, $S_{\mathrm{eo}}$, is mathematically identical.  The bandwidth of the conversion is set by the total mechanical damping $\Gamma_{\mathrm{e}}+\Gamma_{\mathrm{o}}+\kappa_{\mathrm{m}}$.  Conversion efficiency is greatest on mechanical resonance ($\omega=\omega_{\mathrm{m}}$) and when the coupling rates are matched $(\Gamma_{\mathrm{e}}=\Gamma_{\mathrm{o}})$ and exceed any intrinsic mechanical damping $\kappa_{\mathrm{m}}$; in this case the maximum conversion efficiency is $\eta_{\mathrm{e}}\eta_{\mathrm{o}}$.  

Conversion gain $\mathcal{A}$ becomes appreciable when $4\omega_{\mathrm{m}}<\kappa_{\mathrm{e}}, \kappa_{\mathrm{o}}$ (and also depends on detunings $\Delta_{\mathrm{e}}$ and $\Delta_{\mathrm{o}}$; see S.I.).  In this parameter regime, the converter begins to act like a linear, phase-insensitive amplifier, and as such can have an apparent efficiency greater than unity at the cost of adding noise \cite{caves1982}.  While amplification might be beneficial for some applications, ideal frequency conversion requires unit gain ($\mathcal{A}_{\mathrm{e}}=\mathcal{A}_{\mathrm{o}}=1$) so that conversion adds as little noise as possible.

\begin{figure*}
\scalebox{.45}{\includegraphics{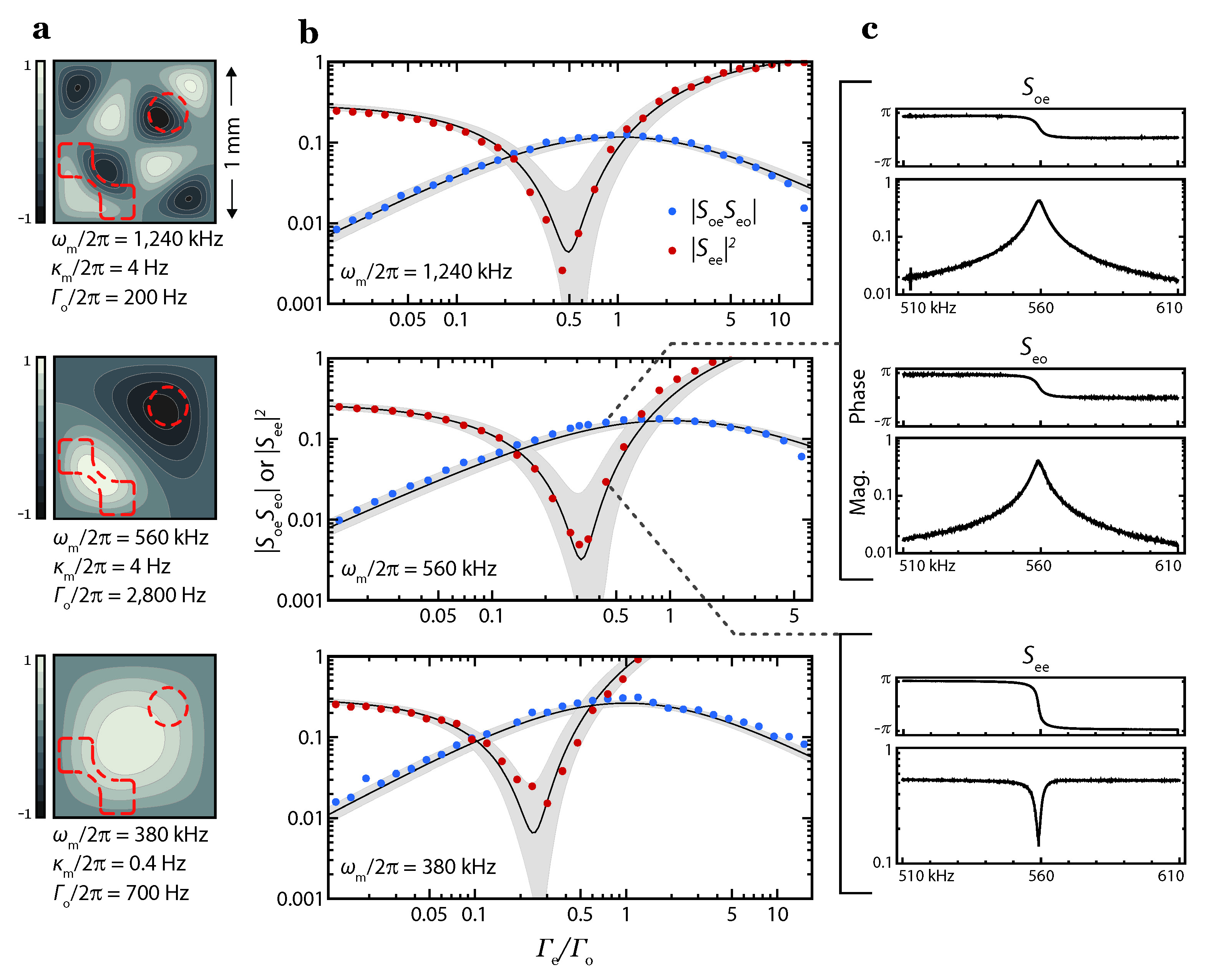}}
\caption{Reversibility and efficiency of conversion. ($\mathbf{a}$) Contour plot of simulated membrane displacement, normalized to unity, for the three vibrational modes used for frequency conversion.  Outlines of the optical mode (dashed circle) and niobium metallization (dashed bowtie) are shown. ($\mathbf{b}$) Apparent transfer efficiency ($|S_{\mathrm{oe}}S_{\mathrm{eo}}|$) and reflected microwave power ($|S_{\mathrm{ee}}|^2$) as a function of damping ratio for the three vibrational modes.  Measurement error is smaller than the plotted points.  The black lines are expected values given independently measured system parameters.  Gray regions express the uncertainty of our system parameters (see S.I.). ($\mathbf{c}$)  $S$-parameters for upconverted ($S_{\mathrm{oe}}$) and downconverted ($S_{\mathrm{eo}}$) signals, and reflected microwave signal ($S_{\mathrm{ee}}$) for the $\omega_{\mathrm{m}}/2\pi=560$ kHz vibrational mode with $\Gamma_{\mathrm{o}}/2\pi=2800$ Hz and $\Gamma_{\mathrm{e}}/2\pi=1300$ Hz.  Frequency is relative to the microwave and optical pump frequencies.  Electromagnetic resonators are centered at 7.1 GHz and 282 THz, with power decay rates $\kappa_{\mathrm{e}}/2\pi=1.6$ MHz and $\kappa_{\mathrm{o}}/2\pi=1.65$ MHz and efficiencies $\eta_{\mathrm{e}}=0.76$ and $\eta_{\mathrm{o}}=0.11$ (0.23 from internal resonator loss and 0.47 from optical modematching).  Pumps are red-detuned with $\Delta_{\mathrm{e}}\approx-\omega_{\mathrm{m}}$ and $\Delta_{\mathrm{o}}/2\pi=-730$ kHz (see S.I.). }
\label{data1}
\end{figure*}

\subsection*{Reversible Frequency Conversion}

By injecting a microwave or optical field and monitoring the outgoing fields, we measure the $S$-parameters that determine the number efficiency of the converter (Fig. \ref{setup}b).  We study conversion using three different vibrational modes of the mechanical resonator, which are shown as membrane-displacement plots in Fig. \ref{data1}a.  Converter performance using the $\omega_{\mathrm{m}}/2\pi=560$ kHz vibrational mode is shown in Fig. \ref{data1}c.  By sweeping the frequency of the injected microwave signal, we confirm that $S_{\mathrm{oe}}$ has the expected Lorentzian lineshape of Eqn. \ref{trans}, with a peak at $\omega=\omega_{\mathrm{m}}$.  Conversion efficiency for the reverse process, $S_{\mathrm{eo}}$, is the same to within our measurement error, giving initial indication that the conversion process is reversible.  We also monitor the reflected microwave power as $S_{\mathrm{ee}}$ during upconversion.  When conversion efficiency peaks at $\omega=\omega_{\mathrm{m}}$, there is a corresponding dip in reflected microwave power, as seen in Fig. \ref{data1}c, further demonstrating that some of the injected microwave signal is indeed being converted.  In addition to the magnitudes of the converted and reflected light, the measurements also yield phase information.  We find that, up to a frequency-dependent phase shift, the phase is preserved during conversion, indicating the conversion between the two frequency regimes is coherent.



While it is difficult to independently and accurately calibrate the quantities $S_{\mathrm{oe}}$ and $S_{\mathrm{eo}}$, we can readily calibrate their product, as diagrammed in Fig. \ref{setup}a.  With this method, we can  determine $S_{\mathrm{oe}}S_{\mathrm{eo}}$ to about 6\% (limited by a combination of standing waves in the microwave transmission lines and sideband folding during direct detection; see S.I.).  We find this calibration procedure appealing because it makes equivalent use of both the forward and reverse transmission of the converter, demonstrating and emphasizing its reversible nature.  The peak value of the apparent conversion efficiency, plotted as $|S_{\mathrm{oe}}S_{\mathrm{eo}}|$, is shown in Fig. \ref{data1}b along with the reflected microwave power, $|S_{\mathrm{ee}}|^2$, at the frequency of peak conversion efficiency.  

The converter performance for all three vibrational modes is correctly predicted by linear optomechanical theory with independently measured system parameters (see S.I.).  By changing the microwave pump power while holding the optical pump power fixed, we vary the ratio $\Gamma_{\mathrm{e}}/\Gamma_{\mathrm{o}}$ and observe that maximum conversion efficiency occurs near $\Gamma_{\mathrm{e}}=\Gamma_{\mathrm{o}}$, as seen in Fig. \ref{data1}b, confirming the impedance matching condition predicted by Eqn. \ref{trans}.  Furthermore, we can explore the effect of conversion gain by using different vibrational modes of the mechanical resonator for frequency conversion.  For the $\omega_{\mathrm{m}}/2\pi=380$ kHz vibrational mode, $\omega_{\mathrm{m}}/\kappa_{\mathrm{e}}\approx \omega_{\mathrm{m}}/\kappa_{\mathrm{o}}\approx 0.25$ and a photon number gain of $\mathcal{A}=3$ is expected.  This results in an apparent efficiency of 0.25, as seen in the peak value of $|S_{\mathrm{oe}}S_{\mathrm{eo}}|$ for this vibrational mode.   For conversion with the higher frequency $\omega_{\mathrm{m}}/2\pi=1.24$ MHz vibrational mode, $\omega_{\mathrm{m}}/\kappa_{\mathrm{e}}\approx\omega_{\mathrm{m}}/\kappa_{\mathrm{o}}\approx0.8$ and the converter approaches the resolved-sideband limit.  This reduces the photon number gain to $\mathcal{A}=1.4$.  With this lower gain comes a lower apparent efficiency, but the absorption of microwave power occurs nearer the peak in conversion efficiency, indicating more ideal conversion (see S.I.).        


The converter reaches nearly unit internal efficiency as $\Gamma_{\mathrm{e}}, \Gamma_{\mathrm{o}}\gg \kappa_{\mathrm{m}}$.  For the $\omega_{\mathrm{m}}/2\pi=560$ kHz vibrational mode, the converter reaches a matched ($\Gamma_{\mathrm{e}}=\Gamma_{\mathrm{o}}$) bandwidth of $30$ kHz, overwhelming the intrinsic mechanical loss by nearly four orders of magnitude; i.e. only about two parts in $10^{4}$ of the converted signal are lost through the mechanical decay channel.  For this vibrational mode, we expect the maximum apparent conversion efficiency to be $|S_{\mathrm{oe}}S_{\mathrm{eo}}|=\mathcal{A}\eta_{e}\eta_{o}=0.169\pm0.012$, which corresponds well with the measured maximum of $0.178\pm0.011$.  The efficiency $\eta_{\mathrm{e}}\eta_{\mathrm{o}}$ does not include conversion gain $\mathcal{A}$, and is limited by absorption and scattering of light in the electromagnetic resonators ($\kappa_{\mathrm{e}, \mathrm{ext}}/\kappa_{\mathrm{e}}=0.76$ and $\kappa_{\mathrm{o}, \mathrm{ext}}/\kappa_{\mathrm{o}}=0.23$), and imperfect optical mode-matching ($\epsilon=0.47$).  These factors combine to give an expected efficiency $\eta_{\mathrm{e}}\eta_{\mathrm{o}}=0.082\pm0.006$.  This corresponds well with the measured maximum apparent efficiency divided by the expected gain of $0.086\pm0.007$.

\subsection*{Vibrational noise}

\begin{figure}
\begin{minipage}{\linewidth}
\scalebox{.42}{\includegraphics{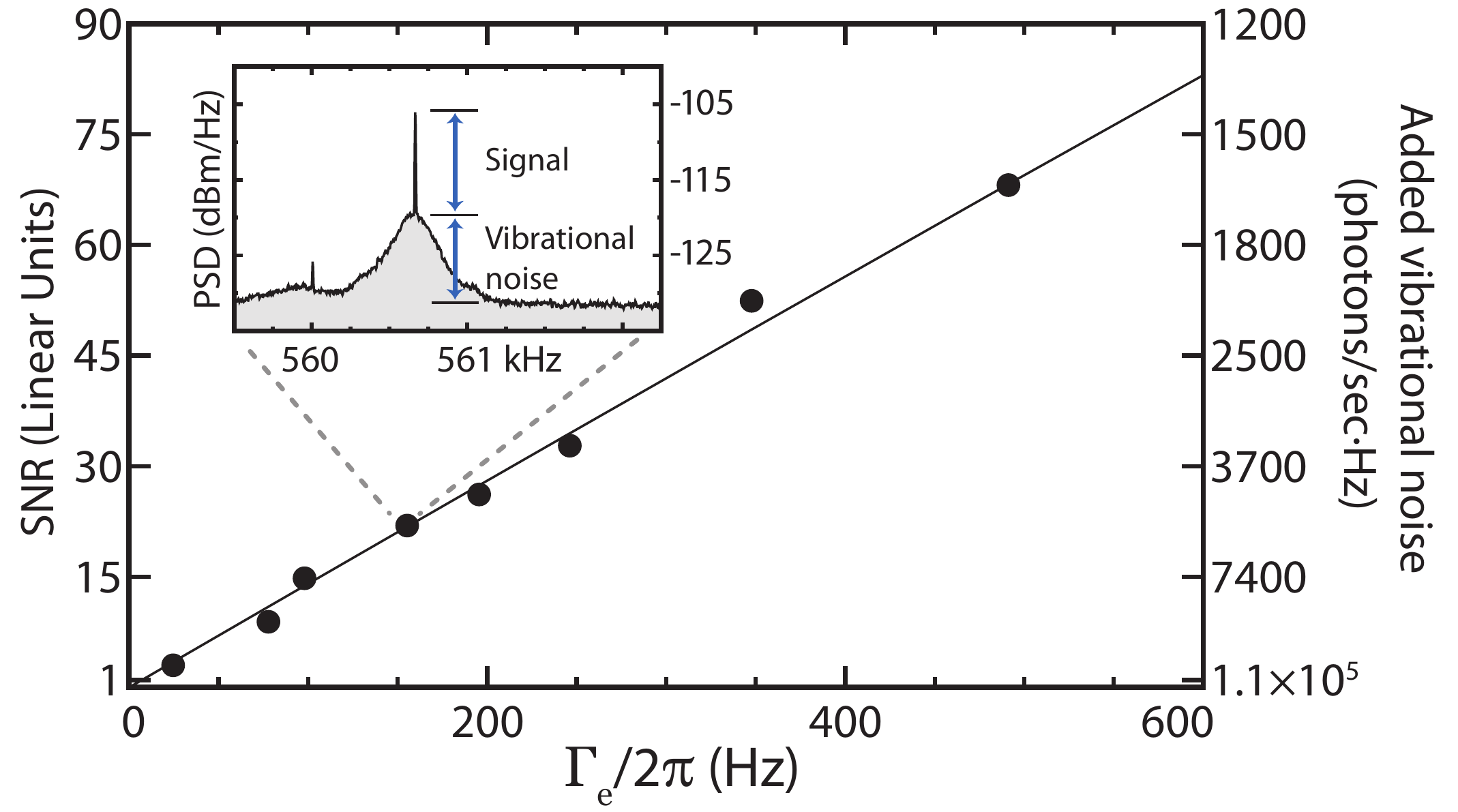}}
\caption{Optically-detected signal to noise ratio (SNR).  SNR for an upconverted microwave signal as a function of coupling rate $\Gamma_{\mathrm{e}}$; injected signal power is kept constant and $\Gamma_{\mathrm{o}}\approx\Gamma_{\mathrm{e}}$.  A fit to optomechanical theory is shown (line), and the extracted added vibrational noise ($n_{\mathrm{vib}}$) is shown on the right vertical axis. Inset: power spectral density (PSD) of an upconverted microwave signal.  }
\label{osnr}
\end{minipage}
\end{figure}

Any extraneous vibrations of the mechanical resonator appear as noise on the converted signal, and can be specified with a signal to noise ratio (SNR), defined as $\mathrm{SNR}=\mathcal{S}_{\mathrm{in}}/n_{\mathrm{vib}}$, where $\mathcal{S}_{\mathrm{in}}$ is the spectral density of the input signal, and $n_{\mathrm{vib}}$ is the spectral density of vibrational noise added during conversion as referenced to the input of the converter \cite{hill2012}.  In units of quanta of added noise (a quantum corresponding to one photon per second in a one hertz bandwidth), upconversion adds
\begin{align}
\label{addednoise}
n_{\mathrm{vib}}=\frac{1}{\mathcal{A}_{\mathrm{e}}\eta_{\mathrm{e}}}\left( \frac{\kappa_{\mathrm{m}}n_{\mathrm{env}}}{\Gamma_{\mathrm{e}}}+ (\mathcal{A}_{\mathrm{e}}-1) + \left(\mathcal{A}_{\mathrm{o}}-1 \right) \right)
\end{align}
for matched coupling rates, where $n_{\mathrm{env}}\equiv 1/(e^{\hbar\omega_{\mathrm{m}}/k_{B}T_{\mathrm{env}}}-1)$, $k_{\mathrm{B}}$ is Boltzmann's constant, and $T_{\mathrm{env}}$ is the temperature of the mechanical resonator's environment.   Any conversion gain adds vibrational noise given by the last two terms in Eqn. \ref{addednoise}.  Mechanical decoherence $\kappa_{\mathrm{m}}n_{\mathrm{env}}$ also adds vibrational noise, but the effect of mechanical decoherence can be reduced by increasing $\Gamma_{\mathrm{e}}$ and $\Gamma_{\mathrm{o}}$.  Only with unit gain ($\eta_{\mathrm{e}}\gg\mathcal{A}_{\mathrm{e}}-1$ and $\eta_{\mathrm{e}}\gg\mathcal{A}_{\mathrm{o}}-1 $) and negligible mechanical decoherence ($\eta_{\mathrm{e}}\Gamma_{\mathrm{e}}\gg\kappa_{\mathrm{m}}n_{\mathrm{env}}$) do vibrations of the mechanical resonator ``freeze out''.  In addition to low vibrational noise, noiseless frequency conversion requires unit conversion efficiency ($\eta_{\mathrm{e}}=\eta_{\mathrm{o}}=1$).


To characterize noise added by vibrations of the mechanical resonator, we inject a fixed-frequency signal into the microwave port of the converter and measure the spectral density of the output optical field.  We measure vibrational noise added during upconversion (as opposed to downconversion) because near quantum-limited photodetection allows us to quickly characterize noise at optical frequencies.  The microwave signal is injected at a frequency $\omega_{\mathrm{m}}$ above the microwave pump with matched coupling rates ($\Gamma_{\mathrm{e}}\approx\Gamma_{\mathrm{o}}$).  As shown in the inset of Fig. \ref{osnr}, the converted signal power appears on a noise background that is dominated by thermal fluctuations filtered by the mechanical resonator's susceptibility.  As we increase the coupling rates $\Gamma_{\mathrm{e}}$ and $\Gamma_{\mathrm{o}}$ (while keeping the injected microwave signal power constant), we observe an increased SNR (Fig. \ref{osnr}).  Fitting the observed SNR to linear optomechanical theory for the spectral density of the injected microwave signal $\mathcal{S}_{\mathrm{in}}$ allows us to estimate $n_{\mathrm{vib}}$, the added vibrational noise.    

Vibrational noise is non-zero during 4 kelvin operation because the mechanical resonator is still ``hot'' and driven by thermal fluctuations.  While this effect can be substantially reduced by increasing $\Gamma_{\mathrm{e}}$ and $\Gamma_{\mathrm{o}}$, colder temperatures are needed to reduce added vibrational noise to the single quantum level.   Thermalizing the vibrational modes of the mechanical resonator to a dilution refrigerator temperature of 40 millikelvin with the achieved $\Gamma_{\mathrm{e}}=\Gamma_{\mathrm{o}}=2\pi\times 15$ kHz would ideally result in $n_{\mathrm{vib}}\approx 1$.  Spurious mechanical modes in the silicon substrate and mirrors that are another potential source of noise are also expected to be suppressed at colder temperatures \cite{purdy2012}.  Furthermore, temperatures of $T_{\mathrm{env}}<( 2\pi\cdot 7 ~\mathrm{GHz})\hbar/k_{\mathrm{B}}\approx$ 300 millikelvin are required to eliminate Johnson noise at microwave frequencies that corrupts weak (quantum) signals.  

\subsection*{Discussion}


We have demonstrated a converter that reversibly, coherently, and efficiently connects the microwave and optical portions of the electromagnetic spectrum.  Three features we have observed---the reversibility of the conversion, its phase-preserving nature, and the absorption of injected signal power during conversion---provide firm evidence that state transfer is occurring, and that we have accessed the beam-splitter Hamiltonian \cite{zhang2003} that is fundamentally capable of noiseless state transfer.  The ability to coherently exchange information between microwave and optical light opens new possibilities for quantum information, in particular entanglement between microwave and optical photons \cite{vitali2012, wang2013,tian2013,kuzyk2013}.

While the converter functions well, some modest changes can significantly improve performance.  Removing internal loss in the electromagnetic resonators will improve $\eta_{\mathrm{e}}$ and $\eta_{\mathrm{o}}$, which currently limit conversion efficiency.  Lower temperatures will improve the microwave resonator, which at 4 kelvin suffers from thermally excited quasiparticles that limit $\eta_{\mathrm{e}}$.  Constructing the optical resonator with asymmetric mirror coatings will improve $\eta_{\mathrm{o}}$ by reducing the amount of light lost through the second mirror.  In addition, more careful placement of the electromechanical system inside the optical resonator can decrease scattered light and improve optical mode matching \cite{purdy2013}.  Added vibrational noise can be reduced by using mechanical resonators with higher quality factors ($Q\equiv\omega_{\mathrm{m}}/\kappa_{\mathrm{m}}$).  In particular, the type of silicon nitride membrane we use is capable of extremely high quality factors of $Q>10^{7}$ \cite{zwickl2008,purdy2012, yu2012}, about two orders of magnitude greater than the quality factors used in this work.  With our modular converter design, improvements to any single converter component can be carried out independently.  This inherent flexibility will allow us to explore the materials and methods necessary to fully integrate optical light with superconducting qubits at millikelvin temperatures.  

 \subsection*{Acknowledgments}

This work was supported by the DARPA QuASAR program and the National Science Foundation under grant number 1125844.  We would like to thank D.R. Schmidt for sharing his knowledge of fabrication techniques, J.N. Ullom for lending us equipment, and P.-L. Yu, J.D. Teufel, and J. Kerckhoff for discussions.  C.A.R. thanks the Clare Boothe Luce foundation for support.  


\bibliography{arxiv-v1-ref}

\onecolumngrid

\newpage

\appendix

\renewcommand\thefigure{\thesection\arabic{figure}}    
\setcounter{figure}{0}    

\renewcommand\thetable{\thesection\arabic{table}}    
\setcounter{table}{0}    

\setcounter{equation}{0}

\setcounter{section}{19}

\section*{Supplementary Information}

\subsection{Electro-opto-mechanical theory}

Our converter is described by a Hamiltonian that includes radiation pressure coupling between electromagnetic and mechanical modes \cite{law1995} :

\begin{align}
H = \hbar \omega_{\text{o}} \hat{a}^\dagger \hat{a} + \hbar \omega_{\text{e}} \hat{b}^\dagger \hat{b} + \hbar \omega_{\text{m}} \hat{c}^\dagger \hat{c} + \hbar G_{\text{o}} (\hat{a}^\dagger \hat{a}) \hat{x} + \hbar G_{\text{e}} (\hat{b}^\dagger \hat{b}) \hat{x} 
\end{align}
Here, $\hat{a}$, $\hat{b}$, and $\hat{c}$ denote annihilation operators for the optical, electrical and mechanical modes, respectively.  Frequencies $\omega_{i}$ describe the mode resonant frequencies, with subscripts $i=\{\mathrm{o,e,m}\}$ denoting optical, electrical, and mechanical.  The couplings from the optical and microwave resonators to the mechanical mode are denoted $G_{\text{o}}$ and $G_{\text{e}}$. Finally, $\hat{x}$ is the mechanical mode position operator $\hat{x}=x_{\mathrm{zp}}(\hat{c}+\hat{c}^\dagger)$ with $x_{\mathrm{zp}}=\sqrt{\frac{\hbar}{2 m \omega_{\text{m}}}}$.

Using this Hamiltonian, we calculate the Heisenberg-Langevin equations of motion \cite{walls2008} and linearize them about a strong, coherent pump field \cite{kippenberg2013}.  We model the optical resonator as having three loss ports, with total linewidth (full-width half-maximum)\footnote{In the main text, we use use $\kappa_{\mathrm{o,L}}\rightarrow\kappa_{\mathrm{o,ext}}$ and $\kappa_{\mathrm{e,R}}\rightarrow\kappa_{\mathrm{e,ext}}$.} $\kappa_{\text{o}} = \kappa_{\text{o,L}} + \kappa_{\text{o,R}} + \kappa_{\text{o,int}}$, comprising the left  mirror (the port that is used during measurement of the converter), right mirror, and internal loss. The microwave resonator has two ports with linewidth $\kappa_{\text{e}} = \kappa_{\text{e,R}} + \kappa_{\text{o,int}}$, and the mechanical resonator has a single port with linewidth $\kappa_{\text{m}}$.  The resulting system of first-order, linear differential equations can be concisely described as a state space model (see, e.g., \cite{kerckhoff2013}). 
\begin{align}
\dot{\mathbf{a}}(t) &=A \mathbf{a}(t) + B \mathbf{a}_{\text{in}}(t)\\
\mathbf{a}_{\text{out}}(t) &= C \mathbf{a}(t) + D \mathbf{a}_{\text{in}}(t)
\end{align}
where $\mathbf{a}=(\hat{a}, \hat{b}, \hat{c}, \hat{a}^\dagger, \hat{b}^\dagger, \hat{c}^\dagger)$ is a vector of the resonator modes that represents the internal state of the converter, and $\mathbf{a}_{\text{in}}$ ($\mathbf{a}_{\text{out}}$) is a vector of input (output) fields. See section \ref{appen} for the form of the matrices $A$, $B$, $C$, and $D$, and the vectors $\mathbf{a}_{\text{in}}$ and $\mathbf{a}_{\text{out}}$. The state-space model can be reduced to a transfer function\footnote{We use $f(\omega)=\int_{-\infty}^{\infty}f(t)e^{-i \omega t}\mathrm{d}t$, which is equivalent to the bilateral Laplace transform $f(s)=\int_{-\infty}^{\infty}f(t)e^{-s t}\mathrm{d}t$ with $s\rightarrow i \omega$.}, $\Xi(\omega)$, that relates input fields $\mathbf{a}_{\text{in}}$ to output fields $\mathbf{a}_{\text{out}}$.
\begin{align}
\mathbf{a}_{\text{out}}(\omega) &= \Xi(\omega) \mathbf{a}_{\text{in}}(\omega)\\
\Xi(\omega) &= C(-i \omega I - A)^{-1}B+D
\end{align}
This formalism reflects our measurement, where we can apply fields (i.e. modify $\mathbf{a}_{\text{in}}$) and measure fields (i.e. measure components of $\mathbf{a}_{\text{out}}$).  Also, many software packages exist to efficiently solve state space and transfer function models.  The S-parameter $S_{\text{oe}}(\omega)$ corresponds to the transfer function element $\Xi_{ a_{\text{out,L}}, b_{\text{in,R}}}(\omega)={a}_{\mathrm{out, L}}(\omega) / {b}_{\text{in,R}}(\omega)$.\footnote{We use $a$ to denote the operator expectation value $\langle \hat{a} \rangle$.}  In the weak-coupling ($\kappa_{\mathrm{e}}, \kappa_{\mathrm{o}}\gg\Gamma_{\mathrm{e}}, \Gamma_{\mathrm{o}}$) limit (which our experiment satisfies), 
\begin{align}
\label{soefull}
S_{\text{oe}}(\omega)= \Xi_{ a_{\text{out,L}}, b_{\text{in,R}}}(\omega) = - \sqrt{\frac{\kappa_{\text{e,R}}}{\kappa_{\text{e}}} \frac{\kappa_{\text{o,L}}}{\kappa_{\text{o}}}\epsilon} \frac{ \sqrt{\Gamma_{\text{e}} \Gamma_{\text{o}}}}{-i(\omega-\omega_{\text{m}}^\prime)+(\Gamma_{\text{e}}+\Gamma_{\text{o}}+\kappa_{\text{m}})/2} \sqrt{\mathcal{A}} \exp ( i\phi )
\end{align}
Here, $\omega_{\text{m}}^\prime = \omega_{\text{m}}+\Omega_{\text{e}}+\Omega_{\text{o}}$, where $\Omega_{\text{e}}$ and $\Omega_{\text{o}}$ are the electro- and optomechanical frequency shifts. $\Gamma_{\text{e}}$ and $\Gamma_{\text{o}}$ are the electro- and optomechanical damping rates (see section \ref{appen}). The gain term $\mathcal{A}$ and the associated phase shift $\phi$ are given by
\begin{align}
\mathcal{A} =\mathcal{A}_\mathrm{e} \times \mathcal{A}_\mathrm{o}= \left[\frac{(\Delta_{\text{e}}-\omega_{\text{m}})^2+(\kappa_{\text{e}}/2)^2}{-4 \Delta_{\text{e}} \omega_{\text{m}}}\right] \times \left[\frac{(\Delta_{\text{o}}-\omega_{\text{m}})^2+(\kappa_{\text{o}}/2)^2}{-4 \Delta_{\text{o}} \omega_{\text{m}}}\right] \\
\phi = \arctan \left(\frac{ \kappa_{\text{e}}/2}{\Delta_{\text{e}}+\omega_{\text{m}}} \right) + \arctan \left(\frac{ \kappa_{\text{o}}/2}{\Delta_{\text{o}}+\omega_{\text{m}}} \right)
\end{align}
For $\Delta_{\text{e}}=\Delta_{\text{o}}=-\omega_{\text{m}}$, $\mathcal{A}$ simplifies to $\left[1+\left(\frac{\kappa_{\text{e}}}{4 \omega_{\text{m}}}\right)^2\right] \left[1+\left(\frac{\kappa_{\text{o}}}{4 \omega_{\text{m}}}\right)^2\right]$, giving it a simple interpretation in terms of the resolved-sideband parameter. We also define the efficiency of the electrical and optical resonators in the main text as $\eta_e=\frac{\kappa_{\text{e,R}}}{\kappa_{\text{e}}}, \eta_o=\epsilon\frac{\kappa_{\text{o,L}}}{\kappa_{\text{o}}}$, where $\epsilon$ is the optical mode matching. In the main text, we omit a constant phase shift.

Reflection measurements are described generically by the following transfer function element:
\begin{align}
\Xi_{a_{\text{out,L}},a_{\text{in,L}}}(\omega) = - \left[1-\frac{\kappa_{\text{o,L}}}{-i(\omega+\Delta_{\text{o}})+\kappa_{\text{o}}/2}\right] \frac{   -i(\omega-\omega_{\text{m}}-\Omega_{\text{e}}-\Omega_{\text{o}} \mathcal{B}) + (\Gamma_{\text{e}}+\Gamma_{\text{o}} \mathcal{B} + \kappa_{\text{m}} )/2 }{-i(\omega-\omega_{\text{m}}-\Omega_{\text{e}}-\Omega_{\text{o}})+(\Gamma_{\text{e}}+\Gamma_{\text{o}}+\kappa_{\text{m}})/2}
\end{align}
where $\mathcal{B}=(1-\frac{i\kappa_{\text{o,L}}/2}{\Delta_{\text{o}}})/(1-\frac{\kappa_{\text{o,L}}}{-i(\omega+\Delta_{\text{o}})+\kappa_{\text{o}}/2})$ is a resonant enhancement factor of order unity. Equivalent expressions for other electromagnetic output fields can be obtained via substitution of appropriate parameters. 

Another path through the system---relevant for signal-to-noise measurements---is from the mechanical bath input to an optical output:
\begin{align}
\Xi_{a_{\text{out,L}},c_{\text{in}}}(\omega) = -i \sqrt{\epsilon\frac{\kappa_{\text{o,L}}}{\kappa_{\text{o}}}}\frac{\sqrt{\Gamma_{\text{o}} \kappa_{\text{m}}}}{-i(\omega-\omega_{\text{m}}^\prime)+(\Gamma_{\text{e}}+\Gamma_{\text{o}}+\kappa_{\text{m}})/2}\sqrt{\mathcal{A}_o} \exp \left[ i \arctan\left(\frac{\kappa_{\text{o}}/2}{\Delta_{\text{o}}+\omega}\right)\right]
\end{align}

Vibrational noise added during conversion can be quantified with the spectral density of the output fields\footnote{We use the notational convention $[\mathbf{a}_{\mathrm{out}}(\omega)]^\dagger=\mathbf{a}_{\text{out}}^\dagger (-\omega)$, as in \cite{clerk2010}.}, $\mathcal{S}(\omega)$,
\begin{align}
2\pi\mathcal{S}(\omega)\delta(\omega-\omega') = \left<  \left[\mathbf{a}_{\text{out}} (\omega')\right]^\dagger \mathbf{a}_{\text{out}}^T(\omega) \right>  =\Xi^\star(\omega') \left< \left[\mathbf{a}_{\text{in}}(\omega')\right]^\dagger \mathbf{a}_{\text{in}}^T(\omega) \right>\Xi^T (\omega)
\end{align}

More concisely, $\mathcal{S}(\omega)=\Xi^\star (\omega) \Sigma \Xi^T (\omega)$, where $\Sigma\equiv (2\pi\delta(\omega-\omega'))^{-1}\left< \left[\mathbf{a}_{\text{in}} (\omega')\right]^\dagger \mathbf{a}_{\text{in}}^T (\omega) \right>$.  Components of the spectral density matrix have units of number/sec$\cdot$Hz.  The covariance matrix $\Sigma$ contains information (see section \ref{appen}) about the bath temperature and quantum noise (e.g. $\left< \left[\hat{c}_{\text{in}} (\omega')\right]^\dagger \hat{c}_{\text{in}}^{}(\omega) \right>=2\pi\delta(\omega-\omega')n_{\mathrm{env}}$ and $\left< \hat{c}_{\text{in}}^{} (\omega) \left[\hat{c}_{\text{in}} (\omega')\right]^\dagger \right>=2\pi\delta(\omega-\omega')(n_{\mathrm{env}}+1)$, where $n_{\mathrm{env}}$ is the mechanical thermal bath phonon occupation \cite{clerk2010}). 

Using this spectral density, we can write a signal to noise ratio (SNR) for upconversion:
\begin{align}
\label{osnrsimple}
\mathrm{SNR} &=\mathcal{S}_{\text{in}}(\omega) \left(   |\Xi_{a_{\text{out,L}},b_{\text{in,R}}}(\omega)|^2 \frac{2 \pi \delta(\omega-\omega^\prime)}{\langle \hat{a}_{\text{out,L}}^\dagger \hat{a}_{\text{out,L}} \rangle}  \right) 
\end{align}
where $\mathcal{S}_{\text{in}}$ is the spectral density of the input microwave field and the second term is noise at the output of the converter referred to the input.  If we assume all fields in $\mathbf{a}_{\mathrm{in}}$ except $\hat{c}_{\mathrm{in}}$ and $\hat{c}^\dagger_{\mathrm{in}}$ are that of vacuum, and additionally use $\Delta_{\text{e}}=\Delta_{\text{o}}=-\omega_{\text{m}}$ and $\Gamma_{\mathrm{e}}=\Gamma_{\mathrm{o}}$, the expression for SNR simplifies to:
\begin{align}
\mathrm{SNR} &= \mathcal{A}_e \frac{\kappa_{\text{e,R}}}{\kappa_{\text{e}}}\left[\frac{\kappa_{\text{m}} n_{\mathrm{env}}}{\Gamma_{\text{e}}}+\left(\frac{\kappa_{\text{e}}}{4 \omega_{\text{m}}}\right)^2 + \left(\frac{\kappa_{\text{o}}}{4 \omega_{\text{m}}}\right)^2 \right]^{-1}  \times \mathcal{S}_{\text{in}}(\omega) \\
 &= \mathcal{A}_e \eta_{\text{e}} \left[\frac{\kappa_{\text{m}} n_{\mathrm{env}}}{\Gamma_{\text{e}}}+\left(\mathcal{A}_{\text{e}}-1\right) + \left(\mathcal{A}_{\text{o}}-1\right) \right]^{-1}  \times \mathcal{S}_{\text{in}}(\omega)
\end{align}
In this case the added vibrational noise, $n_{\mathrm{vib}}$, is found through $\text{SNR} = \mathcal{S}_{\text{in}}/n_\text{vib}$.  This expression accurately describes the measured SNR in Fig. 4 of the main text, even though input microwave fields are not vacuum but a thermal state, because the mechanical bath is the dominant source of noise ($n_{\mathrm{vib}}\gg n_{\mathrm{th,e}}$; see section \ref{appen}).

\subsection{Measurement Network}

\begin{figure}
\begin{minipage}{\linewidth}
\scalebox{.58}{\includegraphics{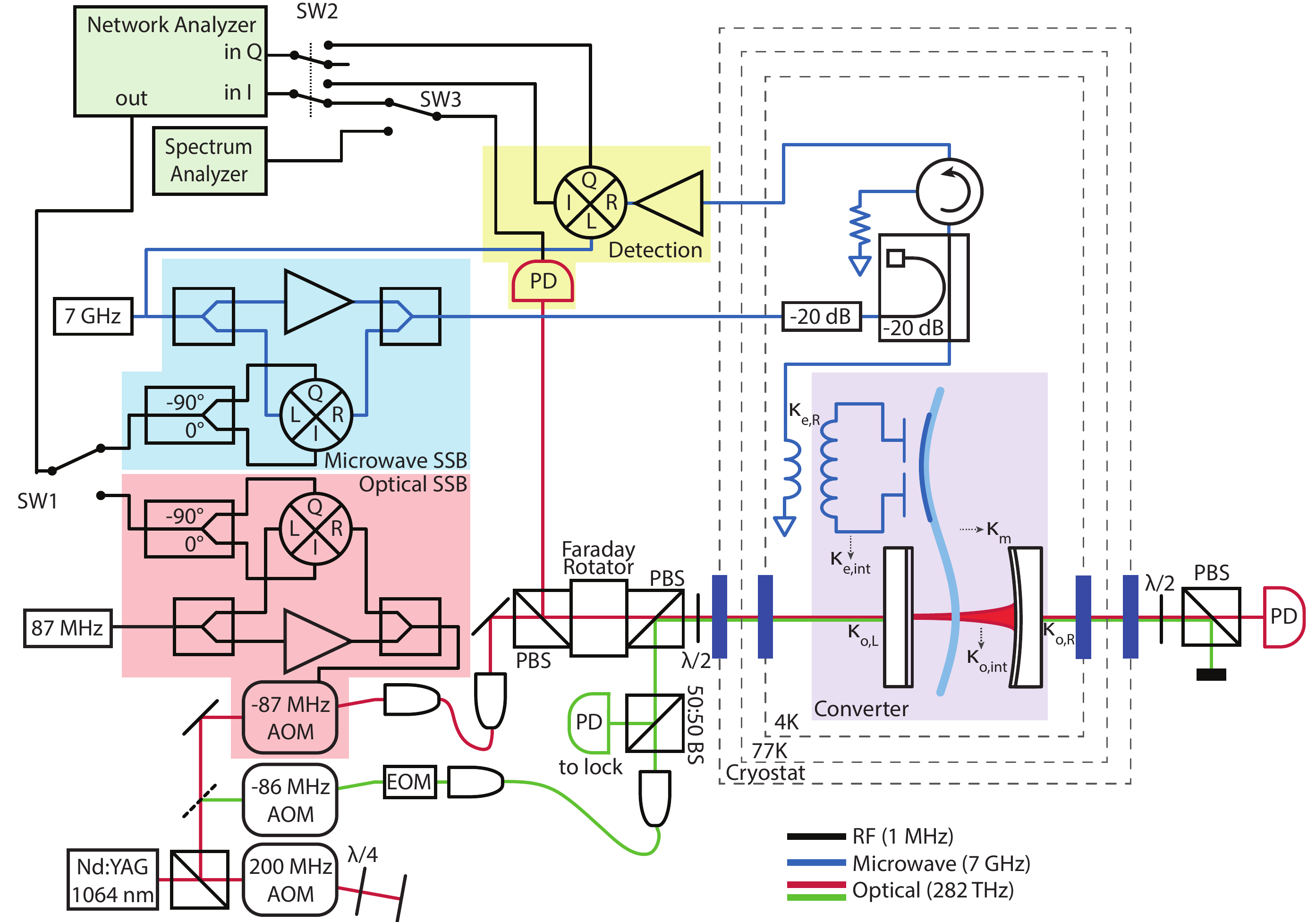}}
\caption{Measurement network.  A network analyzer (green shade) generates a swept signal near the mechanical resonant frequency of interest.  This signal is mixed to a higher frequency using the microwave or optical pump as a reference via single-sideband (SSB) modulation.  The microwave SSB modulation (blue) is made directly on the pump; the optical SSB modulation (pink) is first made on an intermediate radio-frequency signal that drives an acousto-optic modulator (AOM). Any stray lower sideband modulation is suppressed at least 20 dB (10 dB) below the upper sideband modulation for the microwave (optical) carrier. The converter (purple) either reflects or transmits/converts the signal originally produced by the network analyzer.  Reflected and transmitted signal fields are demodulated and detected (yellow) using homodyne (direct) detection for microwave (optical) fields. SNR data is collected by modulating the microwave pump with a signal at frequency $\omega=\omega_{\text{m}}^\prime$ and analyzing the converted optical signal on a spectrum analyzer.  As drawn, the network measures $S_{\text{oe}}$. Throwing switches SW1 and SW2 measures $S_{\text{eo}}$. Throwing SW3 allows signal to noise measurements of the $S_{\text{oe}}$ path. Input/output coupling rates of the converter are noted.}
\label{suppsetup}
\end{minipage}
\end{figure}

Fig. \ref{suppsetup} provides component-level details of the network used to measure performance of the converter.    



\subsubsection{Measurement of converter}

\begin{figure}
\begin{minipage}{\linewidth}
\scalebox{1}{\includegraphics{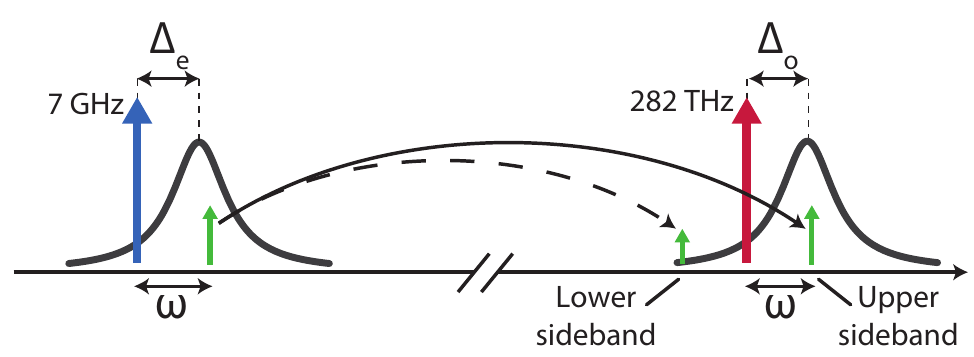}}
\caption{Frequency-domain representation of the conversion process.  A microwave pump (blue arrow) is applied below the microwave resonance (response shown as black curve) with detuning $\Delta_{\mathrm{e}}$.  Likewise, an optical pump (red arrow) is applied below the optical resonance
(response shown as a black curve) with detuning $\Delta_{\mathrm{o}}$.  This allows a signal (green arrow) to be converted.  The conversion also produces a lower sideband when the optical resonator response does not completely suppress it.  Suppression occurs when $4\omega_{\mathrm{m}}\gg\kappa_{\mathrm{o}}$.  }
\label{sb}
\end{minipage}
\end{figure}

For an S-parameter measurement such as $S_{\text{oe}}(\omega)=a_{\text{out,L}}(\omega)/b_{\text{in,R}}(\omega)$, we generate an upper sideband on the microwave pump (with amplitude and phase given by $b_{\text{in,R}}(\omega)$; $b_{\text{in,R}}^\dagger (\omega)$ describes the lower sideband) and measure the upper sideband on the outgoing optical pump (which yields the amplitude and phase of $a_{\text{out,L}}(\omega)$).  In general, the conversion process also produces a non-zero lower sideband; that is, transfer function element that connects $b_{\text{in,R}}(\omega)$ to $a_{out, L}^\dagger (\omega)$ is nonzero (see Fig. \ref{sb}).  In the resolved-sideband limit, the electromagnetic resonators greatly suppress the lower sideband.  In our experiment, with a resolved sideband parameter of $\omega_{\mathrm{m}}/\kappa_{\mathrm{e}}\approx\omega_{\mathrm{m}}/\kappa_{\mathrm{o}}\approx0.5$, we must account for the lower sideband.  During downconversion ($S_{\mathrm{eo}}$) and microwave reflection ($S_{\mathrm{ee}}$), we collect both in-phase and quadrature data in homodyne detection.  No information is lost during measurement, and we can reconstruct the amount of power present in the upper- and lower- sidebands of the reflected microwave pump.  During upconversion ($S_{\mathrm{oe}}$), the reflected optical pump and sidebands are focused onto a photodetector, and we only detect amplitude modulation of the pump.  In order to assign a measured amplitude modulation an equivalent upper-sideband amplitude, we need to know the relationship between the upper- and lower- sidebands.    

The photocurrent we measure is proportional to the incident photon flux (in units of number/sec).  The photon flux is 
\begin{equation}
f(t)\equiv \left<\hat{d}^{\dagger}(t)\hat{d}(t)\right>
\end{equation}
where $\hat{d}(t)$ is the annihilation operator for the light field incident on the detector and has units of $\sqrt{\mathrm{number}/\mathrm{sec}}$. 
During an upconversion measurement, the mode-matched portion of the optical pump, given by $\sqrt{\epsilon}\alpha_{\mathrm{p}}e^{-i\omega_{\mathrm{p}}t}$, is injected into the converter.  The injected pump  induces an intracavity field $\sqrt{ \epsilon}\alpha_{\mathrm{p}}e^{-i\omega_{\mathrm{p}}t}\sqrt{ \kappa_{\mathrm{o, L}}}/(-i\Delta_{\mathrm{o}}+\kappa_{\mathrm{o}}/2)$.   Mechanical motion modulates this intracavity field and generates sidebands at a frequency $\omega_{\text{m}}$ from the pump frequency.  A mechanical displacement $x_{0}\mathrm{sin}(\omega_{\text{m}}t+\Phi)$ results in a total intracavity field of \cite{schliesser2008-2}

\begin{equation}
a(t)=\sqrt{ \epsilon}\alpha_{\mathrm{p}}e^{-i\omega_{\mathrm{p}}t}\frac{\sqrt{ \kappa_{\mathrm{o, L}}}}{-i\Delta_{\mathrm{o}}+\kappa_{\mathrm{o}}/2}\left(1+\frac{G_{\text{o}}x_{0}}{2} \left(\frac{e^{-i(\omega_{\text{m}}t+\Phi)}}{-i(\Delta_{\mathrm{o}}+\omega_{\text{m}})
+\kappa_{\mathrm{o}}/2} -\frac{e^{i(\omega_{\text{m}}t+\Phi)}}{-i(\Delta_{\mathrm{o}}-\omega_{\text{m}})
+\kappa_{\mathrm{o}}/2}\right)\right).
\end{equation}
This results in an outgoing optical field that we detect of 

\begin{align}
\nonumber
d(t)=& \sqrt{ \epsilon}\alpha_{\mathrm{p}}e^{-i\omega_{\mathrm{p}}t}\left(\frac{\kappa_{\mathrm{o, L}}}{-i\Delta_{\mathrm{o}}+\kappa_{\mathrm{o}}/2}-1\right) \\
&+\sqrt{ \epsilon}\alpha_{\mathrm{p}}e^{-i\omega_{\mathrm{p}}t}\frac{\kappa_{\mathrm{o, L}}}{-i\Delta_{\mathrm{o}}+\kappa_{\mathrm{o}}/2} \frac{G_{\text{o}}x_{0}}{2} \left(\frac{e^{-i(\omega_{\text{m}}t+\Phi)}}{-i(\Delta_{\mathrm{o}}+\omega_{\text{m}})
+\kappa_{\mathrm{o}}/2} -\frac{e^{i(\omega_{\text{m}}t+\Phi)}}{-i(\Delta_{\mathrm{o}}-\omega_{\text{m}})
+\kappa_{\mathrm{o}}/2}\right).
\label{detected}
\end{align}

This expression can be interpreted as consisting of the reflected pump field at frequency $\omega_{\mathrm{p}}$ along with an upper sideband at frequency $\omega_{\mathrm{p}}+\omega_{\text{m}}$ and a lower sideband at frequency $\omega_{\mathrm{p}}-\omega_{\text{m}}$.  The lower sideband is negligible in the resolved sideband regime ($\omega_{\text{m}}>\kappa_{\mathrm{o}}$) and when $\Delta_{\mathrm{o}}\approx -\omega_{\text{m}}$.  The non-modematched light in the pump field does not beat with the sidebands on the photodetector, and appears as a zero-frequency photocurrent.  Defining $\theta=\mathrm{Arg}[\kappa_{\mathrm{o, L}}/(-i\Delta_{\mathrm{o}}+\kappa_{\mathrm{o}}/2)-1]$, $\psi=\mathrm{Arg}[\kappa_{\mathrm{o, L}}/(-i\Delta_{\mathrm{o}}+\kappa_{\mathrm{o}}/2)]$, and neglecting the zero-frequency (DC) portion of the photocurrent, 

\begin{align}
f(t)=\mathcal{G}&e^{-i(\omega_{\text{m}}t+\Phi)}   
  \left(1-e^{2i(\theta-\psi)}\left(   \frac{-i(\Delta_{\mathrm{o}}+\omega_{\text{m}})+\kappa_{\mathrm{o}}/2}{i(\Delta_{\mathrm{o}}-\omega_{\text{m}})+\kappa_{\mathrm{o}}/2}  \right)\right)+\mathrm{c.c.}\\
\label{flux1}
\nonumber
\mathcal{G}&=\epsilon\left|\alpha_{\mathrm{p}}\right|^{2}\left|\frac{\kappa_{\mathrm{o, L}}}{-i\Delta_{\mathrm{o}}+\kappa_{\mathrm{o}}/2}-1\right|  \left|\frac{\kappa_{\mathrm{o, L}}}{-i\Delta_{\mathrm{o}}+\kappa_{\mathrm{o}}/2}\right|  \frac{G_{\text{o}}x_{0}}{2} \left( \frac{e^{-i(\theta-\psi)}}{-i(\Delta_{\mathrm{o}}+\omega_{\text{m}})+\kappa_{\mathrm{o}}/2}\right)   
\end{align}

In a measurement where we do not detect the lower sideband at frequency $\omega_{\mathrm{p}}-\omega_{\text{m}}$, we would detect a photocurrent proportional to $f_{ssb}(t)$.   
\begin{equation}
\frac{f_{\mathrm{ssb}}(t)}{f(t)}=\frac{1}{1-e^{2i(\theta-\psi)}\left(   \frac{-i(\Delta_{\mathrm{o}}+\omega_{\text{m}})+\kappa_{\mathrm{o}}/2}{i(\Delta_{\mathrm{o}}-\omega_{\text{m}})+\kappa_{\mathrm{o}}/2}    \right)}
\end{equation}
This ratio allows us to convert between the measured photon flux $f(t)$ and an equivalent upper-sideband photon flux $f_{\mathrm{ssb}}(t)$.  

As a note, the pump is mildly attenuated by the optical cavity (the factor $ \kappa_{\mathrm{o, L}}/(-i\Delta_{\mathrm{o}}+\kappa_{\mathrm{o}}/2)-1 $ in Eqn. \ref{detected}).  This changes the effective gain seen by the sidebands during photodetection.  During calibration (discussed in the main text), we account for this effect.      

\subsubsection{Optical network}

The optical network utilizes two laser beams: A pump tone (Fig. \ref{suppsetup}, red line) and a locking tone (Fig. \ref{suppsetup}, green line), both of which are derived from a single 1064 nm Nd:YAG laser (Coherent, Inc. Mephisto).  Both beams are double-passed through a 200 MHz AOM that is used to lock the laser frequency to the optical resonator.  The locking tone provides feedback to the 200 MHz AOM via the Pound-Drever-Hall (PDH) stabilization technique.  The locking tone operates at low powers and near optical cavity resonance to minimize unwanted optomechanical effects.  The pump tone is detuned 1 MHz red from the locking tone (accomplished by passing each through different $\sim$80 MHz AOMs with a relative detuning of 1 MHz) and occupies a polarization mode orthogonal to the locking beam, resulting in a detuning of $\Delta_{\mathrm{o}}/2\pi=-730$ kHz.  The birefringent splitting of the cavity is 270 kHz.  The AOM used to detune the pump from the locking tone is also used to write the SSB signal used in $S$-parameter measurements onto the pump.  A Faraday rotator, polarizing beamsplitters (PBSs), and a half-wave plate inject the tones into orthogonal polarization modes of the optical cavity, and separate the reflected light for detection with photodetectors. Transmitted light is also detected and used to measure optomechanical parameters.  

\subsubsection{Optical calibration}

\begin{table}
\caption{\label{parameters}Converter parameters. Sets of values apply to the three mechanical modes of interest.}
\begin{ruledtabular}
\begin{tabular}{c  c  c  c  c }
$\kappa_{\text{e}}/2\pi$ (kHz)	&	$\kappa_{\text{e,R}}/2\pi$ (kHz)	&	$\kappa_{\text{e,int}}/2\pi$ (kHz)	&	$\Delta_{\text{e}}/2\pi$ (kHz)	&	$G_{\text{e}} x_{\mathrm{zp}}/2\pi$ (Hz)\\
1590$\pm$30	&	1220$\pm$30	&	370$\pm$20	&	\{-520, -650, -1350\}$\pm$50	&	\{2.1$\pm$0.1, 2.7$\pm$0.1, 0.76$\pm$0.04\}	\\
$\kappa_{\text{o}}/2\pi$ (kHz)	&	$\kappa_{\text{o,L}}/2\pi$ (kHz)	&	$\kappa_{\text{o,int}}/2\pi$ (kHz)	&	$\Delta_{\text{o}}/2\pi$ (kHz)	&	$G_{\text{o}} x_{\mathrm{zp}}/2\pi$ (Hz)	\\
1650$\pm$50	&	380$\pm$70	&	850$\pm$120	&	-730$\pm$40	&	\{7$\pm$1 , 23$\pm$4, 15$\pm$3\}	\\
& $\kappa_{\text{o,R}}/2\pi$ (kHz) & & $\omega_{\text{m}}/2\pi$ (kHz)	&	$\kappa_{\text{m}}/2\pi$ (Hz)    \\
& 420$\pm$80 & &\{380, 560, 1240\} & \{0.42$\pm$0.05, 4.0$\pm$0.5, 4.2$\pm$0.5\}  
\end{tabular}
\end{ruledtabular}
\end{table}

A summary of optomechanical parameters is in Table \ref{parameters}. In general, calibration is performed as described in previous work \cite{purdy2012}. The optical coupling $G_{\text{o}}x_{\text{zp}}$ is extracted from a measurement of mechanical damping, using the thermal motion of the membrane at temperature $T_{\text{env}}$ as a known displacement standard. This differs from the electromechanical calibration of $G_{\text{e}}x_{\text{zp}}$ in that this method does not require separate knowledge of the intracavity power.

We directly measure the output coupling rate $\kappa_{\text{o,L}}$ from peak transmission and reflection measurements of the cavity in both directions \cite{hood2001}. This measurement has five unknown quantities: the mirror transmissions $T_{\text{L}}$ and $T_{\text{R}}$, the internal loss $L$, and the mode matching factors $\epsilon_{\text{L}}$, and $\epsilon_{\text{R}}$. (Here, $\epsilon_{\text{L}}$ is the mode matching $\epsilon$ used in the rest of the text.) Peak power measurements of the cavity in the forward direction (left port input) are $P_{\text{in,f}}$, $P_{\text{r,f}}$, and $P_{\text{t,f}}$, indicating input, reflection, and transmission. We can write the cavity transmission and reflection as
\begin{align}
\label{cal1}
\frac{P_{\text{t,f}}}{\epsilon_{\text{L}} P_{\text{in,f}}} & = 4 T_{\text{L}} T_{\text{R}} \left( \frac{1}{T_{\text{L}}+T_{\text{R}}+L} \right)\\
\label{cal2}
\frac{\epsilon_{\text{L}} P_{\text{r,f}}-(1-\epsilon_{\text{L}})P_{\text{in,f}}}{\epsilon_{\text{L}} P_{\text{in,f}}} & = \left(\frac{L + T_{\text{R}} - T_{\text{L}}}{L+T_{\text{R}}+T_{\text{L}}}\right)^2
\end{align}
We can eliminate the modematching $\epsilon_{\text{L}}$ by dividing Eqn. \ref{cal1} by Eqn. \ref{cal2}:
\begin{align}
\frac{P_{\text{t,f}}}{P_{\text{r,f}}-P_{\text{in,f}}} = \frac{4 T_{\text{L}} T_{\text{R}}}{(L+T_{\text{R}}-T_{\text{L}})^2-(L+T_{\text{R}}+T_{\text{L}})^2}
\end{align}
This expression now contains only directly-measurable optical powers. This and an equivalent expression for probing the cavity from the opposite side, combined with a measurement of the cavity finesse $\mathcal{F}=\frac{2\pi}{T_{\text{L}}+T_{\text{R}}+L}$, specify $T_{\text{L}}$, $T_{\text{R}}$, and $L$ uniquely. We can convert to the loss rate of interest via $\kappa_{\text{o,L}} = \kappa_{\text{o}} \frac{T_{\text{L}}}{T_{\text{L}}+T_{\text{R}}+L}$, likewise for $\kappa_{\text{o,R}}$, $\kappa_{\text{o,int}}$. Once these are determined, we can solve for the modematching $\epsilon_{\text{L}}$ and $\epsilon_{\text{R}}$.


\subsubsection{Microwave network}

The electrical network utilizes a microwave signal generator (Agilent Technologies) operating at $\sim$7 GHz.  This source provides the microwave pump tone, and provides the local oscillator for two quadrature-IF mixers (Marki Microwave) that are used to modulate and demodulate the pump tone with the signal used in $S$-parameter measurements.  The microwave cavity is intrinsically stable (environmental fluctuations change the microwave resonant frequency by much less than its linewidth), and no stabilization technique is needed.

\subsubsection{Microwave calibration}

\begin{figure}
\begin{minipage}{\linewidth}
\scalebox{.70}{\includegraphics{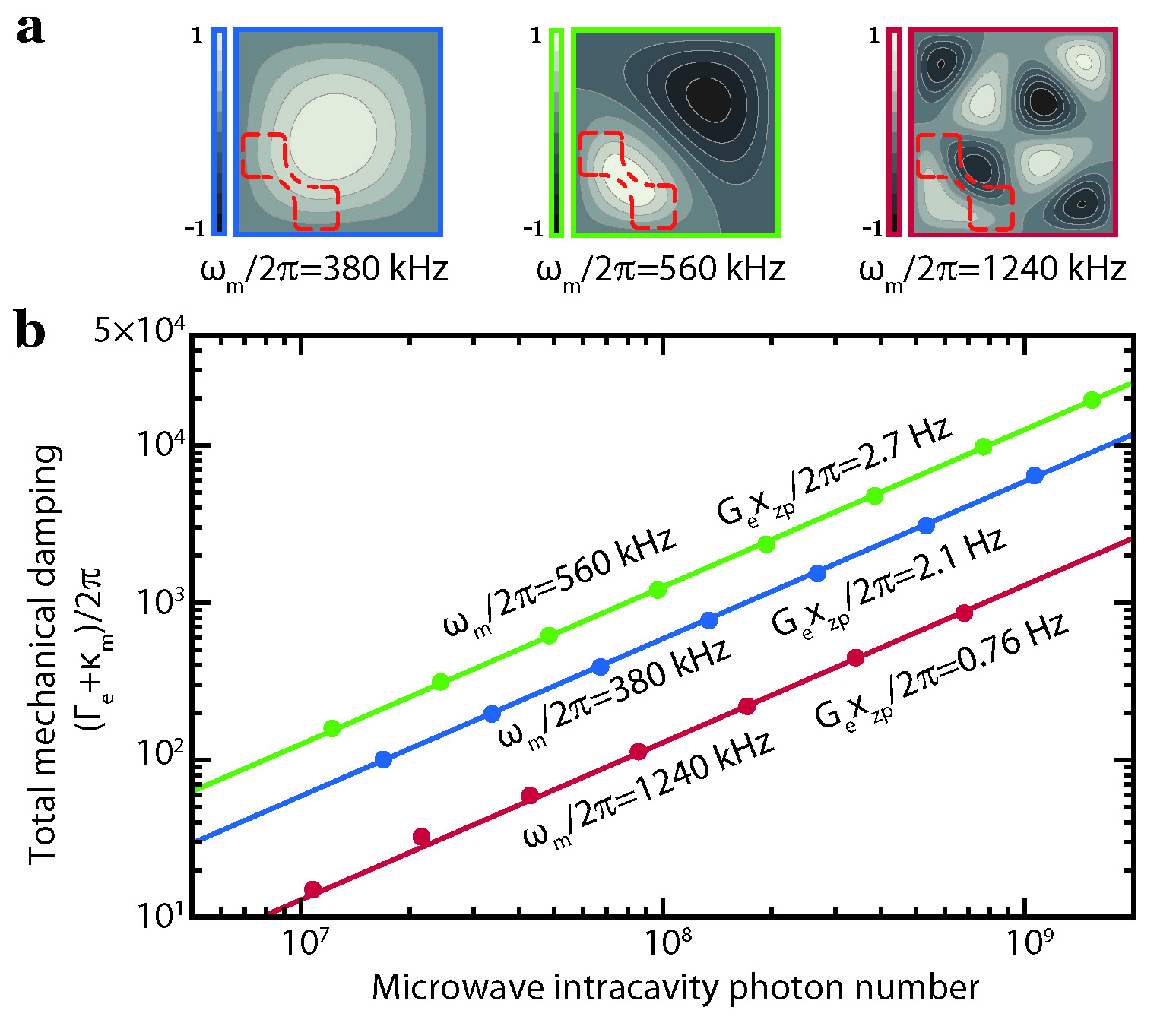}}
\caption{($\mathbf{a}$) Contour plot of membrane displacement for three different vibrational modes.  ($\mathbf{b}$) Mechanical damping for these vibrational modes as a function of microwave pump power.  The electromechanical coupling rates, $G_{\text{e}}x_{\text{zp}}$, are found by fitting the data to linear optomechanical theory.}
\label{eitcal}
\end{minipage}
\end{figure}

Parameters of the microwave resonator, detailed in Table \ref{parameters}, are found by sweeping a signal over a broad range of frequencies that includes the central frequency of the microwave resonator.  Fitting the reflected signal allows us to extract the external coupling rate ($\kappa_{\text{e,R}}$) and the total cavity linewidth ($\kappa_{e}$).  The electromechanical coupling rate is measured by monitoring the total mechanical damping as a function of the strength of the microwave pump.  Mechanical damping is inferred from the mechanically dressed response of the microwave resonator, sometimes referred to as optomechanically induced transparency \cite{weis2010}.  Mechanical damping rate as a function of microwave pump power for the three different mechanical modes used in this work is shown in Fig. \ref{eitcal}. 

\subsection{Additional data}

\begin{figure}
\begin{minipage}{\linewidth}
\scalebox{.36}{\includegraphics{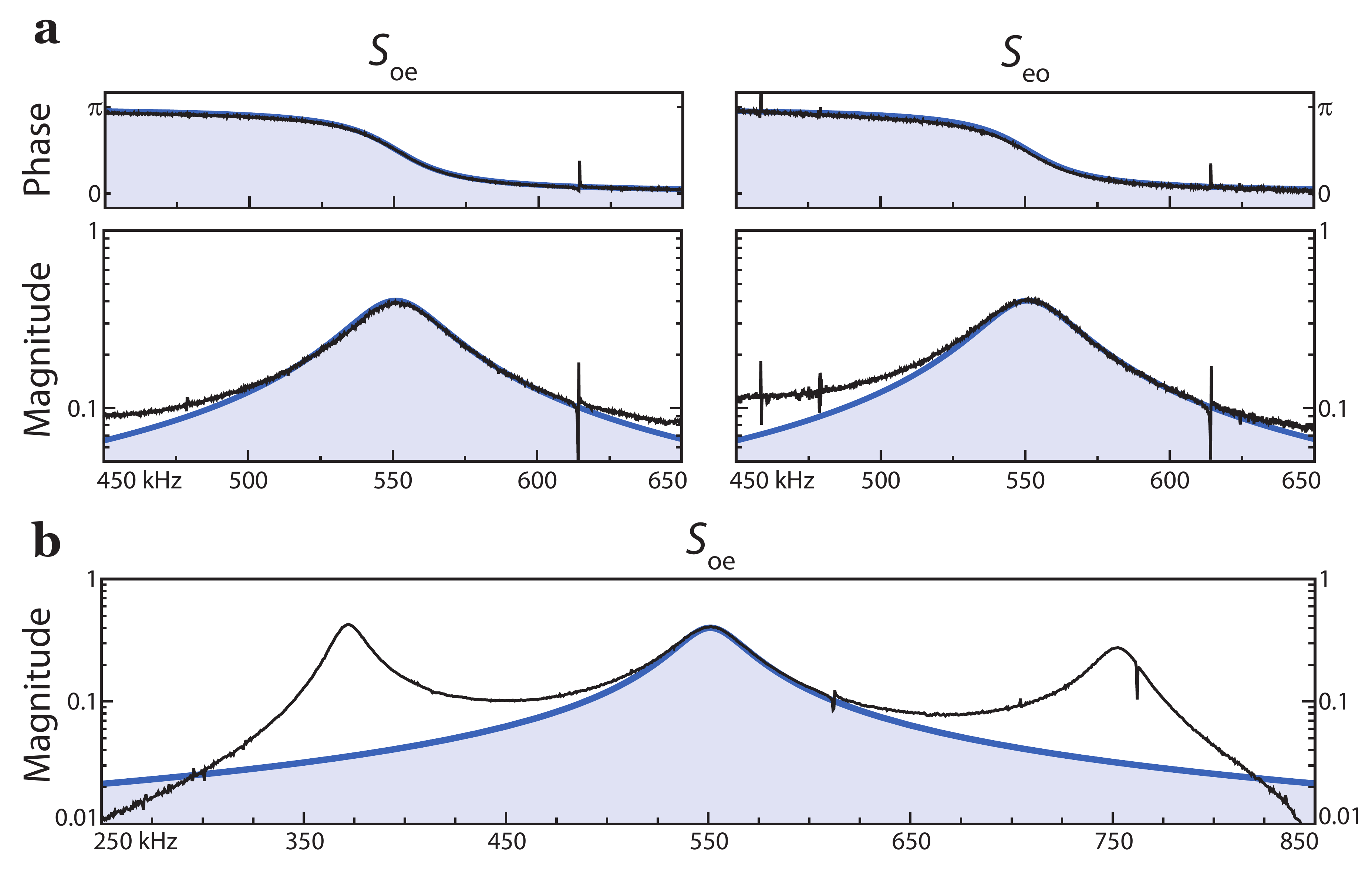}}
\caption{($\mathbf{a}$) $S$-parameters for upconversion ($S_{\mathrm{oe}}$) and downconversion ($S_{\mathrm{eo}}$) for the $\omega_{\mathrm{m}}/2\pi=560$ kHz vibrational mode with 32 kHz of transfer bandwidth.  A Lorentzian fit (blue) is included.  ($\mathbf{b}$) Upconversion over a wide range of frequencies; conversion is visible for three vibrational modes of the mechanical resonator at $\omega_{\mathrm{m}}/2\pi=380$ kHz, 560 kHz, and 760 kHz.}
\label{wideband}
\end{minipage}
\end{figure}

The converter reaches a matched $(\Gamma_{\mathrm{e}}=\Gamma_{\mathrm{o}}$) bandwidth of 32 kHz, as shown in Fig. \ref{wideband}a.   With large bandwidths, measurable conversion occurs across a wide range of frequencies.  In particular, the deviation from a Lorentzian frequency dependence during conversion seen in Fig. \ref{wideband}a can be partially explained by the presence of nearby mechanical modes shown in Fig. \ref{wideband}b.  The optomechanical theory we use is only appropriate when the vibrational modes of the mechanical resonator are spectrally resolved.    

\subsection{Compatibility of superconducting circuitry and optical frequency light}

\begin{figure}
\begin{minipage}{\linewidth}
\scalebox{.60}{\includegraphics{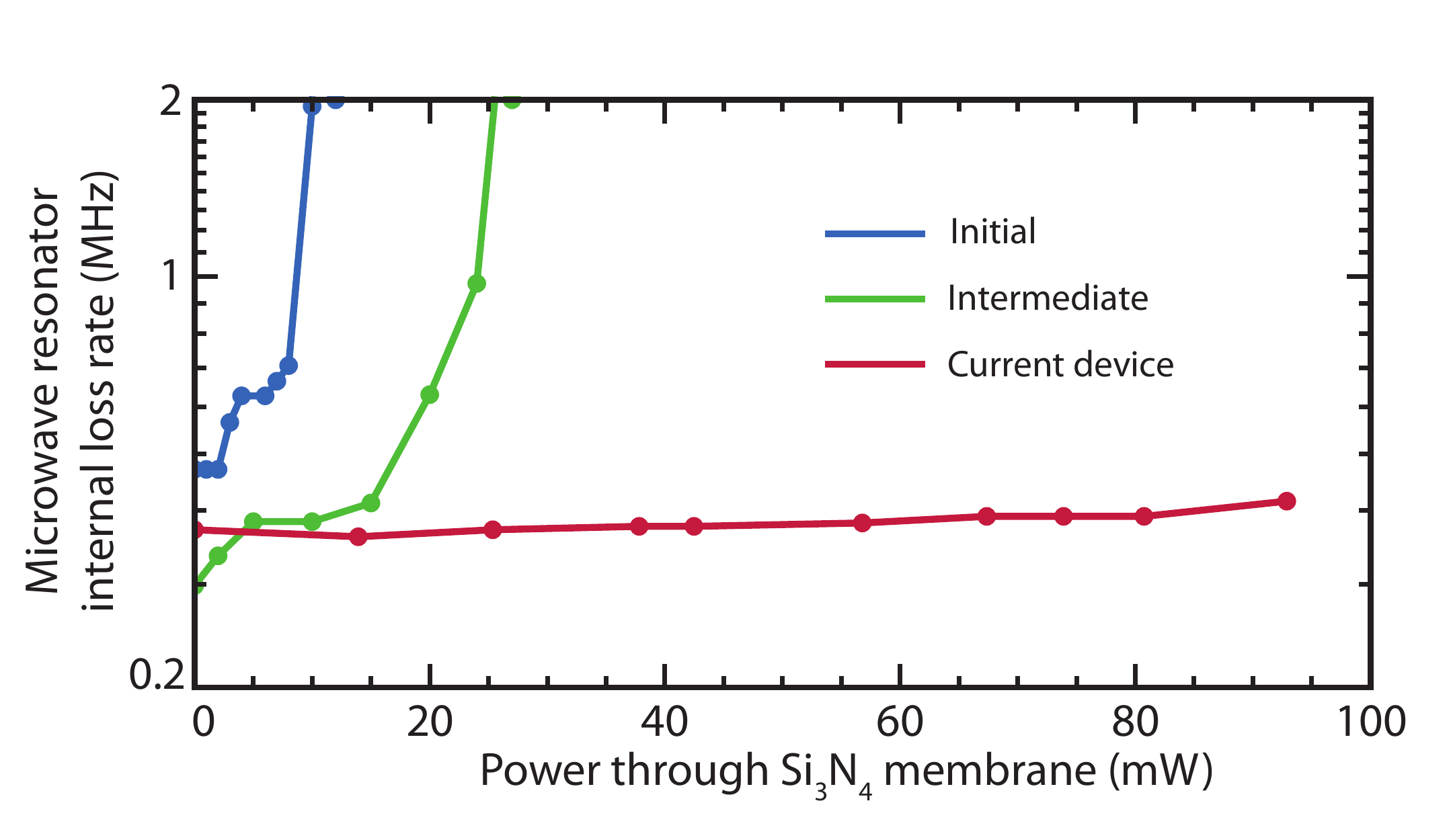}}
\caption{Internal microwave resonator loss rate, $\kappa_{\text{e,int}}$, as a function of optical power.  Initial converters only functioned with low ($<$10 mW) optical powers. } 
\label{pwrhand}
\end{minipage}
\end{figure}

The large circulating optical powers ($\sim$100 mW) necessary for converter operation could adversely affect cryogenic performance in two ways.  First, the silicon nitride membrane that forms the mechanical resonator can absorb a small fraction of the optical photons that pass through it, resulting in localized heating that possibly increases mechanical decoherence. Based on previous work, we expect that the mechanical resonator does not absorb enough optical power to significantly increase the $\sim$4 kelvin bath temperature seen by the mechanical resonator \cite{purdy2013}; furthermore, membrane heating would affect the niobium metallization on the membrane, resulting in additional observed loss in the microwave circuitry (particularly evident if local temperatures exceeded 9 kelvin, the superconducting transition temperature of niobium).  Second, stray light absorbed by the niobium can directly break Cooper pairs and create additional quasiparticles and loss with no need for bulk heating.  We have observed this behavior.  As shown in Fig. \ref{pwrhand}, circulating optical powers of ~10 mW greatly increased internal loss in the microwave resonator for early tests of this type of microwave-to-optical converter.  For the device used in this work, additional shielding was added to prevent stray light in the optical resonator from reaching the niobium microwave circuitry.  With this shielding, the maximum optomechanical coupling ($\Gamma_{\mathrm{o}}$) can reach a similar value to the maximum electromechanical coupling ($\Gamma_{\mathrm{e}}$) for the $\omega_{\mathrm{m}}/2\pi=560$ kHz and $\omega_{\mathrm{m}}/2\pi=380$ kHz vibrational modes.    


\subsection{Appendix}
\label{appen}

In the expressions presented here, $g_{\text{e}} = G_{\text{e}} x_{\text{zp}}$ and $g_{\text{o}} = G_{\text{o}} x_{\text{zp}}$ are electro- and optomechanical couplings in Hz. The electrical pump produces a classical intracavity field described by $\beta$, where $|\beta|=\sqrt{n_{\text{e}}}$ and $\mathrm{Arg}[\beta] = \theta_{\text{p}}$.  Likewise, the optical pump produces a classical intracavity field described by $\alpha$, where $|\alpha|=\sqrt{n_{\text{o}}}$ and $\mathrm{Arg}[\alpha] = \phi_{\text{p}}$.  We also use $n_{\text{th,m}}=n_{\text{env}}$ to describe the mechanical bath phonon occupation, and use $n_{\text{th,e}}$ ($n_{\text{th,o}}$) to describe the average photon occupation of the electrical (optical) input fields at temperature.
\begin{align}
\Omega_{\text{e}} = g_{\text{e}}^2 \beta^2\left[ \frac{\Delta_{\text{e}}+\omega}{(\Delta_{\text{e}}+\omega)^2+(\kappa_{\text{e}}/2)^2}+\frac{\Delta_{\text{e}}-\omega}{(\Delta_{\text{e}}-\omega)^2+(\kappa_{\text{e}}/2)^2}\right] \\
\Omega_{\text{o}} = g_{\text{o}}^2 \alpha^2\left[ \frac{\Delta_{\text{o}}+\omega}{(\Delta_{\text{o}}+\omega)^2+(\kappa_{\text{o}}/2)^2}+\frac{\Delta_{\text{o}}-\omega}{(\Delta_{\text{o}}-\omega)^2+(\kappa_{\text{o}}/2)^2}\right] \\
\Gamma_{\text{e}} = g_{\text{e}}^2 \beta^2  \left[ \frac{\kappa_{\text{e}}}{(\Delta_{\text{e}}+\omega)^2+(\kappa_{\text{e}}/2)^2} -  \frac{\kappa_{\text{e}}}{(\Delta_{\text{e}}-\omega)^2+(\kappa_{\text{e}}/2)^2} \right] \\
\Gamma_{\text{o}} = g_{\text{o}}^2 \alpha^2  \left[ \frac{\kappa_{\text{o}}}{(\Delta_{\text{o}}+\omega)^2+(\kappa_{\text{o}}/2)^2} -  \frac{\kappa_{\text{o}}}{(\Delta_{\text{o}}-\omega)^2+(\kappa_{\text{o}}/2)^2} \right]
\end{align}

\begin{align}
A &= \left(
\begin{array}{cccccc}
 i \text{$\Delta_{\text{o}} $}-\frac{\text{$\kappa_{\text{o}} $}}{2} & 0 & -i e^{i \phi_{\text{p}} } g_{\text{o}} \alpha  & 0 & 0 & -i e^{i \phi_{\text{p}} } g_{\text{o}} \alpha  \\
 0 & i \text{$\Delta_{\text{e}}$}-\frac{\text{$\kappa_{\text{e}} $}}{2} & -i e^{i \theta_{\text{p}} } g_{\text{e}} \beta  & 0 & 0 & -i e^{i \theta_{\text{p}} } g_{\text{e}} \beta  \\
 -i e^{-i \phi_{\text{p}} } g_{\text{o}} \alpha  & -i e^{-i \theta_{\text{p}} } g_{\text{e}} \beta  & -\frac{\text{$\kappa_{\text{m}} $}}{2}-i \text{$\omega_{\text{m}} $} & -i e^{i \phi_{\text{p}} } g_{\text{o}} \alpha  & -i e^{i \theta_{\text{p}} } g_{\text{e}} \beta  & 0 \\
 0 & 0 & i e^{-i \phi_{\text{p}} } g_{\text{o}} \alpha  & -i \text{$\Delta_{\text{o}} $}-\frac{\text{$\kappa_{\text{o}} $}}{2} & 0 & i e^{-i \phi_{\text{p}} } g_{\text{o}} \alpha  \\
 0 & 0 & i e^{-i \theta_{\text{p}} } g_{\text{e}} \beta  & 0 & -i \text{$\Delta_{\text{e}} $}-\frac{\text{$\kappa_{\text{e}} $}}{2} & i e^{-i \theta_{\text{p}} } g_{\text{e}} \beta  \\
 i e^{-i \phi_{\text{p}} } g_{\text{o}} \alpha  & i e^{-i \theta_{\text{p}} } g_{\text{e}} \beta  & 0 & i e^{i \phi_{\text{p}} } g_{\text{o}} \alpha  & i e^{i \theta_{\text{p}} } g_{\text{e}} \beta  & i \text{$\omega_{\text{m}} $}-\frac{\text{$\kappa_{\text{m}} $}}{2} \\
\end{array}
\right)
\end{align}

\begin{align}
B &= \left(
\begin{array}{cc}
M & 0 \\
0 & M \\
\end{array}
\right)
\end{align}

\begin{align}
M&=\left(
\begin{array}{cccccc}
\sqrt{\text{$\kappa_{\text{o,L}} $}} & \sqrt{\text{$\kappa_{\text{o,R}} $}} & \sqrt{\text{$\kappa_{\text{o,int}} $}} & 0 & 0 & 0 \\
 0 & 0 & 0 & \sqrt{\text{$\kappa_{\text{e,R}} $}} & \sqrt{\text{$\kappa_{\text{e,int}} $}} & 0 \\
 0 & 0 & 0 & 0 & 0 & \sqrt{\text{$\kappa_{\text{m}} $}} \\
\end{array}
\right)
\end{align}

\begin{align}
C &= \left(
\begin{array}{cccccc}
 \sqrt{\text{$\kappa_{\text{o,L}} $}} & 0 & 0 & 0 & 0 & 0 \\
 \sqrt{\text{$\kappa_{\text{o,R}} $}} & 0 & 0 & 0 & 0 & 0 \\
 0 & \sqrt{\text{$\kappa_{\text{e,R}} $}} & 0 & 0 & 0 & 0 \\
 0 & 0 & 0 & \sqrt{\text{$\kappa_{\text{o,L}} $}} & 0 & 0 \\
 0 & 0 & 0 & \sqrt{\text{$\kappa_{\text{o,R}} $}} & 0 & 0 \\
 0 & 0 & 0 & 0 & \sqrt{\text{$\kappa_{\text{e,R}} $}} & 0 \\
\end{array}
\right)
\end{align}

\begin{align}
D &= \left(
\begin{array}{cccccccccccc}
 -1 & 0 & 0 & 0 & 0 & 0 & 0 & 0 & 0 & 0 & 0 & 0 \\
 0 & -1 & 0 & 0 & 0 & 0 & 0 & 0 & 0 & 0 & 0 & 0 \\
 0 & 0 & 0 & -1 & 0 & 0 & 0 & 0 & 0 & 0 & 0 & 0 \\
 0 & 0 & 0 & 0 & 0 & 0 & -1 & 0 & 0 & 0 & 0 & 0 \\
 0 & 0 & 0 & 0 & 0 & 0 & 0 & -1 & 0 & 0 & 0 & 0 \\
 0 & 0 & 0 & 0 & 0 & 0 & 0 & 0 & 0 & -1 & 0 & 0 \\
\end{array}
\right)
\end{align}

\begin{align}
\Sigma &= \left(
\begin{array}{cc}
N & 0 \\
0 & N + I \\
\end{array}
\right)
\end{align}

\begin{align}
N &= \left(
\begin{array}{cccccc}
 n_{\text{th,o}} & 0 & 0 & 0 & 0 & 0 \\
 0 & n_{\text{th,o}} & 0 & 0 & 0 & 0 \\
 0 & 0 & n_{\text{th,o}} & 0 & 0 & 0 \\
 0 & 0 & 0 & n_{\text{th,e}} & 0 & 0 \\
 0 & 0 & 0 & 0 & n_{\text{th,e}} & 0 \\
 0 & 0 & 0 & 0 & 0 & n_{\text{th,m}}  \\
\end{array}
\right)
\end{align}

\begin{align}
\mathbf{a}_{\text{in}} &= \left(
\begin{array}{cccccccccccc}
 \hat{a}_{\text{in,L}} & \hat{a}_{\text{in,R}} & \hat{a}_{\text{in,int}} & \hat{b}_{\text{in,R}} & \hat{b}_{\text{in,int}} & \hat{c}_{\text{in}} & \hat{a}_{\text{in,L}}^\dagger & \hat{a}_{\text{in,R}}^\dagger & \hat{a}_{\text{in,int}}^\dagger & \hat{b}_{\text{in,R}}^\dagger & \hat{b}_{\text{in,int}}^\dagger & \hat{c}_{\text{in}}^\dagger \\
\end{array}
\right)^T\\
\mathbf{a}_{\text{out}} &= \left(
\begin{array}{cccccc}
\hat{a}_{\text{out,L}} & \hat{a}_{\text{out,R}} & \hat{b}_{\text{out,R}} & \hat{a}_{\text{out,L}}^\dagger & \hat{a}_{\text{out,R}}^\dagger & \hat{b}_{\text{out,R}}^\dagger\\
\end{array}
\right)^T
\end{align}


\end{document}